\documentclass[fleqn,usenatbib]{mnras}
\usepackage{float}
\usepackage[utf8]{inputenc}
\DeclareUnicodeCharacter{223C}{\ensuremath{}}
\DeclareUnicodeCharacter{2212}{\ensuremath{}}
\DeclareUnicodeCharacter{0301}{\ensuremath{}}

\usepackage{newtxtext,newtxmath}
\usepackage[T1]{fontenc}

\DeclareRobustCommand{\VAN}[3]{#2}
\let\VANthebibliography\thebibliography
\def\thebibliography{\DeclareRobustCommand{\VAN}[3]{##3}\VANthebibliography}

\usepackage{graphicx}	% Including figure files
\usepackage{amsmath}	% Advanced maths commands
\usepackage{caption}
\usepackage{lscape}
\title[Deep ASKAP EMU Survey of the G23 field]{Deep ASKAP EMU Survey of the GAMA23 field: Properties of radio sources}

\author[G. G\"urkan et al.]{
G\"ulay G\"urkan,$^{1,2}$\thanks{E-mail: gulaygurkan.astro@gmail.com}
I. Prandoni,$^{3}$
A. O'Brien,$^{15,5,4}$
W. Raja,$^{5}$
L. Marchetti,$^{6,3}$
M. Vaccari,$^{7,3}$
\newauthor
S. Driver,$^{8}$
E. Taylor,$^{9}$
T. Franzen,$^{10}$
M. J. I. Brown,$^{11}$
S. Shabala,$^{12}$
H. Andernach,$^{13}$
A. M. Hopkins,$^{14}$ 
\newauthor
R. P. Norris,$^{5,15}$
D. Leahy,$^{16}$
M. Bilicki,$^{17}$
H. Farajollahi,$^{15}$
T. Galvin,$^{18}$
G. Heald,$^{2}$
\newauthor
B. S. Koribalski,$^{5,15}$
T. An,$^{19,20}$
K. Warhurst$^{2}$
\\
$^{1}$ Th\"uringer Landessternwarte, Sternwarte 5, D-07778 Tautenburg, Germany\\
$^{2}$ CSIRO Space and Astronomy, ATNF, PO Box 1130, Bentley WA 6102, Australia\\
$^{3}$ INAF-IRA, Via P. Gobetti 101, I-40129 Bologna, Italy\\
$^{4}$ Department of Physics, University of Wisconsin-Milwaukee, P.O. Box 413, Milwaukee, WI 53201, USA\\
$^{5}$ CSIRO Space and Astronomy, ATNF, P.O. Box 76, Epping, NSW 1710, Australia\\
$^{6}$ Department of Astronomy, University of Cape Town, 7701 Rondebosch, Cape Town, South Africa\\
$^{7}$ Inter-university Institute for Data Intensive Astronomy, Department of Astronomy,
University of Cape Town, 7701 Rondebosch, Cape Town, South Africa\\
$^{8}$  International Centre for Radio Astronomy Research/University of Western 
Australia, Perth, Australia\\
$^{9}$ Centre for Astrophysics and Supercomputing, Swinburne University of Technology, Hawthorn 3122, Australia\\
$^{10}$ ASTRON: the Netherlands Institute for Radio Astronomy, PO Box 2, 7990 AA, Dwingeloo, The Netherlands\\
$^{11}$ School of Physics and Astronomy, Monash University, Clayton, Victoria 3800, Australia\\
$^{12}$ School of Natural Sciences, Private Bag 37, University of Tasmania, Hobart, TAS 7001, Australia\\
$^{13}$ Depto. de Astronom\'{i}a, DCNE, Universidad de Guanajuato, Callej\'on. de Jalisco s/n, Guanajuato, CP 36023, Mexico\\
$^{14}$ Australian Astronomical Optics, Macquarie University, 105 Delhi Rd,
North Ryde, NSW 2113, Australia\\
$^{15}$ Western Sydney University, Locked Bag 1797, Penrith, NSW 2751, Australia\\
$^{16}$ Dept. Physics and Astronomy, University of Calgary, Calgary, Alberta, Canada T2N 1N4\\
$^{17}$ Center for Theoretical Physics, Polish Academy of Sciences, al. Lotników 32/46, 02-668 Warsaw, Poland\\
$^{18}$ International Centre for Radio Astronomy Research - Curtin University, 1 Turner Avenue, Bentley WA 6102, Australia\\
$^{19}$ Shanghai Astronomical Observatory, Chinese Academy of Sciences, Nandan road 80,
Shanghai, 200030, China\\
$^{20}$ Key Laboratory of Cognitive Radio and Information Processing, Guilin University
of Electronic Technology, 541004 Guilin, China\\}
\date{Accepted XXX. Received YYY; in original form ZZZ}

\pubyear{2015}

\begin{document}
\label{firstpage}
\pagerange{\pageref{firstpage}--\pageref{lastpage}}
\maketitle

\begin{abstract}
We present the Australian Square Kilometre Array Pathfinder (ASKAP) observations of the Galaxy and Mass Assembly (GAMA)-23h field. The survey was carried out at 887.5 MHz and covers a $\sim$83 square degree field. We imaged the calibrated visibility data, taken as part of the Evolutionary Mapping of Universe (EMU) Early Science Programme, using the latest version of the $\texttt{ASKAPSoft}$ pipeline. The final mosaic has an angular resolution of 10 arcsec and a central rms noise of around 38 $\mu$Jy beam$^{-1}$. The derived radio source catalogue has 39812 entries above a peak flux density threshold of 5$\sigma$. We searched for the radio source host galaxy counterparts using the GAMA spectroscopic (with an i-band magnitude limit of 19.2 mag) and multi-wavelength catalogues that are available as part of the collaboration. We identified hosts with GAMA spectroscopic redshifts for 5934 radio sources. We describe the data reduction, imaging, and source identification process, and present the source counts. Thanks to the wide area covered by our survey, we obtain very robust counts down to 0.2 mJy. ASKAP’s exceptional survey speed, providing efficient, sensitive and high resolution mapping of large regions of the sky in conjunction with the multi-wavelength data available for the GAMA23 field, allowed us to discover 63 giant radio galaxies. The data presented here demonstrate the excellent capabilities of ASKAP in the pre-SKA era.

\end{abstract}

\begin{keywords}
surveys -- catalogs -- radio continuum: general -- radio continuum: galaxies
\end{keywords}

%%%%%%%%%%%%%%%%%%%%%%%%%%%%%%%%%%%%%%%%%%%%%%%%%%

%%%%%%%%%%%%%%%%% BODY OF PAPER %%%%%%%%%%%%%%%%%%

\section{Introduction}
%See \texttt{mnras\_sample.tex} for a more complex example, 
%\citet or citep{deLaguarde1903, delaGuarde1904}.
In the extragalactic Universe there are two main mechanisms producing non-thermal synchrotron emission: (i) relativistic magnetised plasma jets that are generated by the accretion of matter onto supermassive black holes (SMBHs) at the centre of jetted (or radio-loud) active galactic nuclei (AGN) and (ii) cosmic ray electrons accelerated by supernova explosions, the end products of massive stars. The latter radio emission can be used as a probe of the recent number of massive stars and therefore as a proxy for the star formation rate  \citep[SFR,][]{condon92}. There are other means for estimating SFRs of galaxies such as ultraviolet (UV) and infrared (IR) luminosity. UV measures are strongly affected by uncertainties due to dust obscuration corrections whereas IR surveys are limited by poor resolution and source blending. Unlike the UV and IR, the radio emission from normal galaxies traces both obscured and unobscured star formation and is free of cirrus contamination that affects IR estimates at faint luminosities \citep{condon92}. In the absence of deep X-ray observations, the luminosity (as well as morphology, spectrum, and polarisation) of radio emission from jetted AGN provides us the only way of assessing the kinetic luminosity produced by the AGN \citep[i.e. jet power; e.g.][]{willott99}, even though large uncertainties remain to be addressed \citep{kdta97,blundellrawlings2000,manalakou02,hk13,turner18,hard18}. AGN radio luminosity is generally expected to be much higher than the radio luminosity produced by star formation, at least for powerful jetted AGN. In low luminosity AGN (at around P<10$^{24}$ W Hz$^{-1}$) both processes might contribute to the radio emission at similar levels \citep[e.g.][]{gurkan18,gurkan19,macfarlane21,mckean21,hartley21}. Hence, the separation of these processes contributing to the total radio emission becomes a challenge. Sub-arcsec resolution radio data would assist us to separate these two processes, but obtaining this for significantly large samples is still challenging, though there is progress toward overcoming this \citep[e.g.][]{morabito21}. Alternatively, one should rely on optical spectroscopy and/or multi-wavelength photometric data. Wide-field, sensitive and high-resolution radio surveys in regions of the sky where optical spectroscopy and dense multi-band data are available will be crucial for comprehending the evolution of AGN \citep{best14,smol17,ocran21}, the link between AGN and their host galaxies \citep[e.g.][]{gurkan15,webster21,mingo22}, the properties and evolution of star-forming galaxies \citep[e.g.][]{gurkan18,djb21} and systematic selection and investigation of rare radio sources \citep[e.g.][]{gurkan21}.

Waiting for the Square Kilometre Array (SKA\footnote{\url{https://www.skatelescope.org/}}) era, pathfinders and precursors of the SKA already expand the frontiers of astrophysical research by performing wide-field radio surveys. At the low end of the radio spectrum (50-300 MHz) recent radio surveys such as the Low Frequency Array \citep[LOFAR;][]{lofar13} Multifrequency Snapshot Sky Survey \citep[MSSS;][]{msss15}, the LOFAR Low-band antenna Sky Survey \citep[LoLSS;][]{lolls2021}, the LOFAR Two-metre Sky Survey \citep[LoTSS;][]{shimwell17,shimwell22}, the GaLactic and Extragalactic All-Sky Murchison Wide-Field Array \citep[MWA;][]{mwa13} Survey \citep[GLEAM;][]{gleam17} and the TIFR Giant Metrewave Radio Telescope \citep[GMRT;][]{swarup91} 150 MHz all-Sky radio Survey \citep[TGSS;][]{tgss17} have proven the impact of sensitive low-frequency radio observations in both galactic and extragalactic science. In the mid-frequency range (800 MHz - 2 GHz) successful wide-field radio surveys include   the Sydney University Molonglo Sky Survey \citep[SUMSS;][]{sumss03}, the 1.4 GHz National Radio Astronomy Observatory (NRAO) Very Large Array (VLA) Sky Survey \citep[NVSS;][]{nvss98}, the Faint Images of the Radio Sky at Twenty Centimeters \citep[FIRST;][]{first95}, the Karl G. Jansky Very Large Array Sky Survey \citep[VLASS;][]{vlass20}.

As a survey instrument the Australian SKA Pathfinder \citep[ASKAP;][]{johnston07,mcc16,hotan21} is capable of surveying wide sky fields rapidly with high resolution and sensitivity. ASKAP is located at the Murchison Radio-astronomy Observatory (MRO) in Western Australia, and is operated by the Commonwealth Scientific and Industrial Research Organisation (CSIRO). A full description of the array is provided by \citet{hotan21}. Here we only mention that ASKAP is an array of 36 12-metre-diameter prime-focus antennas distributed over a 6 km-diameter region; each is equipped with an innovative phased-array feed \citep[PAF;][]{hay06} that enables the simultaneous digital formation of 36 dual-polarisation beams to sample a field of view of up to 31 square degrees. ASKAP can observe in the frequency range of 800-1800 MHz with an instantaneous bandwidth of 288 MHz. A shallow (about 250 net hours) ASKAP survey at 887.5 MHz \citep[RACS;][]{mcconnell20,hale21} has already completed its first phase. There is another ongoing radio continuum survey with ASKAP  the "Evolutionary Map of Universe" \citep[EMU;][]{norris11} to provide a deeper radio view of the full Southern radio sky and up to +30 degrees in the Northern sky. EMU started its early science program in 2018 and reached around 25 $\mu$Jy beam$^{-1}$ root-mean-square (rms) noise at around 12 arcsec resolution \citep[e.g.][]{joseph19,pennock21,norris21}.

The Galaxy and Mass Assembly \citep[GAMA\footnote{\url{http://www.gama-survey.org/}};][]{driver16} spectroscopic survey has collected spectra for $\sim$300,000 galaxies down to r$<$19.8 mag over $\sim$286 deg$^{2}$. It was carried out using the AAOmega multi-object spectrograph on the Anglo-Australian Telescope (AAT) by the GAMA team. One of the fields observed as part of GAMA is a 50 deg$^{2}$ region, centred at the right ascension $\alpha$J2000 $=$ 23 hour and the declination $\delta$J2000 $=-32^{\circ}$ and is referred to as GAMA23 (or G23). In G23, spectroscopy is limited to objects brighter than i-band$=$19.2 mag. 

G23 was observed as part of ASKAP commissioning \citep{leahy19} and, later, as part of the EMU Early Science program. This region has also been observed by the far-IR (100-500$\mu$m) instrument Herschel Space Observatory \citep[H-ATLAS;][]{smith17} and a few other key surveys such as Wide-field Infrared Survey Explorer \citep[WISE;][]{wise10}, Galaxy Evolution Explorer \citep[GALEX;][]{galex03}, the Kilo-Degree Survey \citep[KiDS;][]{dejong15} and the Visible and Infrared Survey Telescope for Astronomy (VISTA) Kilo-degree Infrared Galaxy \citep[VIKING;][]{edge13} survey. There is also a dedicated Australian Telescope Compact Array (ATCA) survey (GAMA Legacy ATCA Southern Survey - GLASS) carried out at 5.5 GHz and 9.5 GHz providing radio data at much higher resolutions \citep[e.g.][]{seymour20}. %The ASKAP commissioning observations over G23 have been published \citep{leahy19}. 
The field has also been targeted by the First Large Absorption Survey in HI (FLASH) early science programme \citep{allison16} and provided upper limits on the HI column density frequency distribution function \citep{allison20}. The current data sets in conjunction with the imminent future surveys such as the Widefield ASKAP L-band Legacy All-sky Blind survey \citep[WALLABY;][]{koribalski20}, the Polarization Sky Survey of the Universe's Magnetism survey \citep[POSSUM;][]{gaensler10},  Deep Investigation of Neutral Gas Origins (DINGO\footnote{\url{https://dingo-survey.org/}}), the Middle Ages Galaxy Properties with Integral Field Spectroscopy programme \citep[MAGPI;][]{foster21} and the upcoming observations with extended ROentgen Survey with an Imaging Telescope Array \citep[eROSITA; e.g. ][]{merloni20,brunner21}, all covering the G23 field, will open further new windows and foster our understanding of the Universe.  

In this paper we describe the observations and re-imaging of EMU Early Science observations of the G23 field (Section \ref{sec:2}).  Assessment of the data quality is presented in Section \ref{sec:3}. In Section \ref{sec:4}, we describe our final data products and the sample properties, focusing on source counts and giant radio galaxies. Our conclusions and future work are given in Section \ref{sec:5}. Throughout the paper we use the most recent Planck cosmology \citep{planck20}: $H_{0}$=67.7 km s$^{-1}$ Mpc$^{-1}$, $\Omega_{m}$=0.308 and $\Omega_{\Lambda}$=0.692. Spectral index $\alpha$ is defined in the sense $S \sim \nu^{-\alpha}$.

\begin{table}
	\centering
	\caption{Observation parameters.}
	\label{tab:params}
	\begin{tabular}{lc} % four columns, alignment for each
		\hline
		\hline
		Frequency & 887.5 MHz\\
		Bandwidth & 288 MHz\\
		Integration per tile & 11 hr\\
		Footprint & square$\_$6x6\\
		Beam (pitch) spacing & 0.9 deg\\
		Area covered & 82.7 deg$^2$\\
		Angular resolution & 10 arcsec\\
		\hline
	\end{tabular}
\end{table}

\section{Observations, data reduction and imaging}\label{sec:2}
\begin{figure}
	\hspace{-3em}\includegraphics[width=110mm]{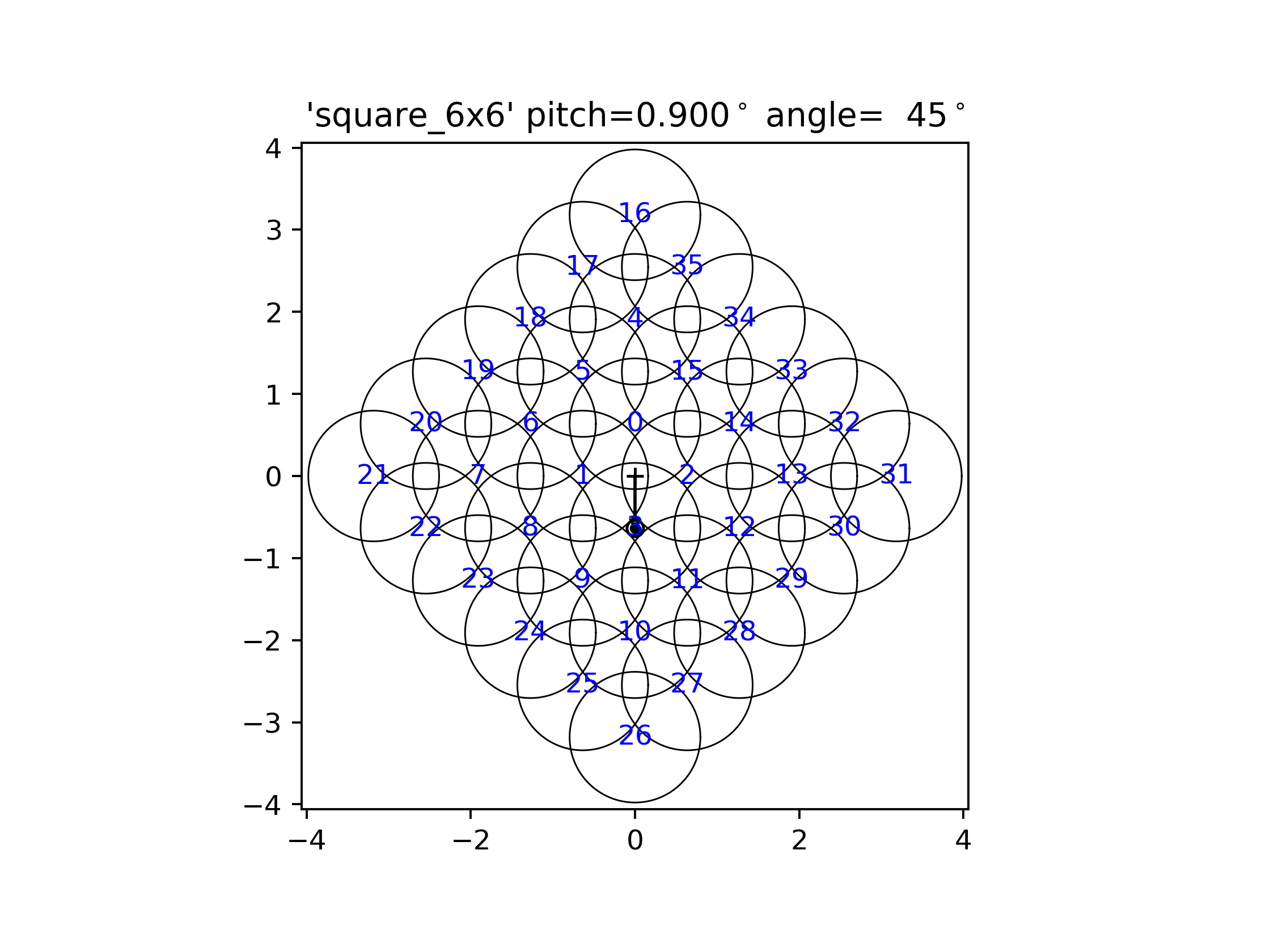}
    \caption{The arrangement of the 36 ASKAP beams in the "square$\_$6x6" configuration. The beams are numbered from 0 to 35.}
    \label{fig:tiles}
\end{figure}

ASKAP observations (Scheduling block identification number: 8132 and 8137) of the G23 field were carried out using the parameters given in Table \ref{tab:params} in March, 2019 as part of the EMU Early Science project. During the observations 35 out of 36 antennas were operating. Observations were carried out using a rectangle tiling configuration. Each tile consists of 36 beams which were arranged as 6 by 6 with a beam spacing (pitch angle) of 0.9 deg. In order to cover the entire G23 region 2 overlapping tiles (with a separation of 0.9 deg) were used (see Fig. \ref{fig:tiles} for design of a single tile). 

\begin{figure*}
	\includegraphics[width=170mm]{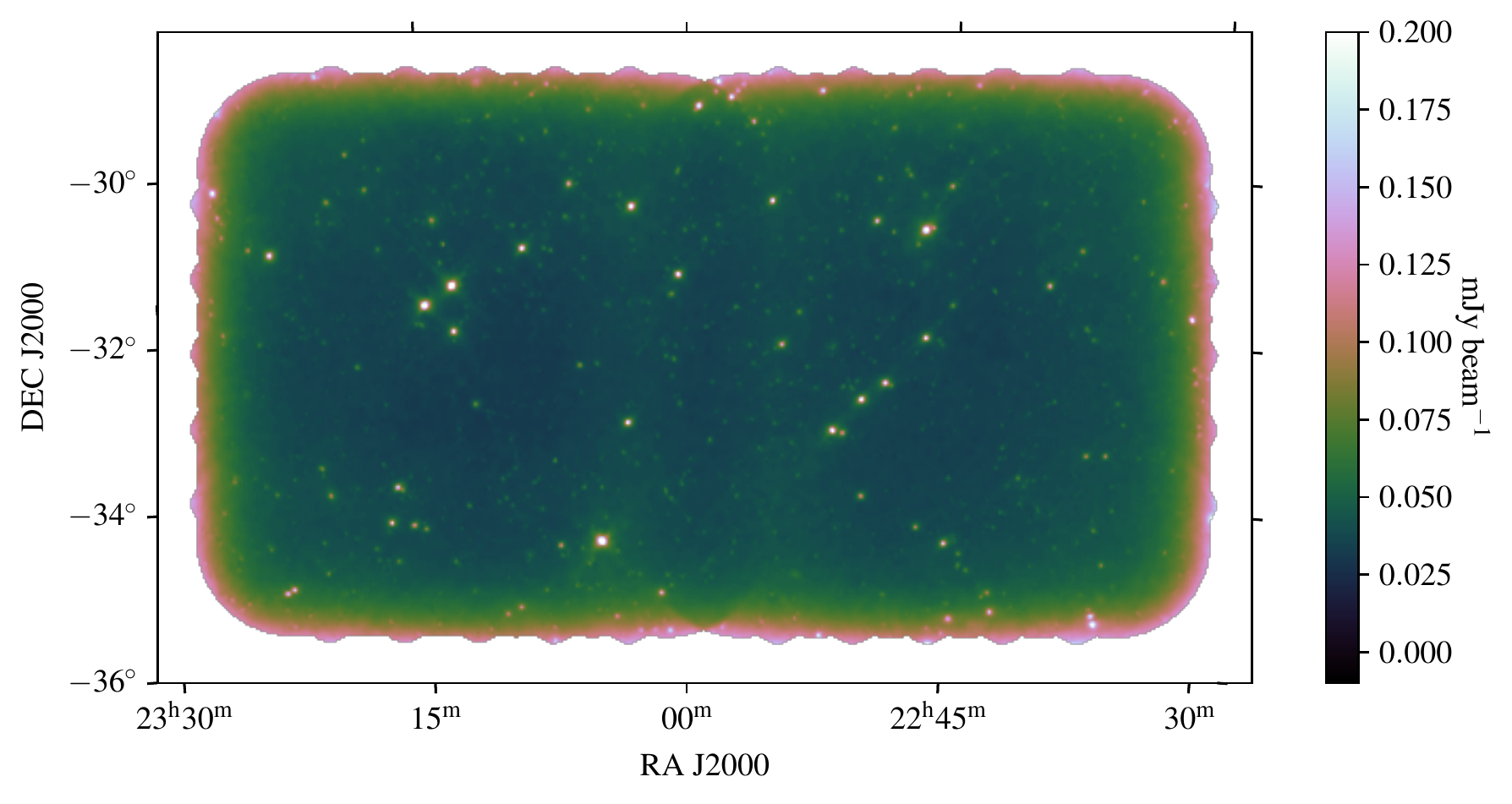}
    \caption{The rms noise map of the G23 ASKAP field. The noise map is produced as part of the source extraction process (see Section \ref{extract} for more details).}
    \label{fig:rms}
\end{figure*}

\begin{figure}
	\hspace{-1.5em}\includegraphics[width=90mm]{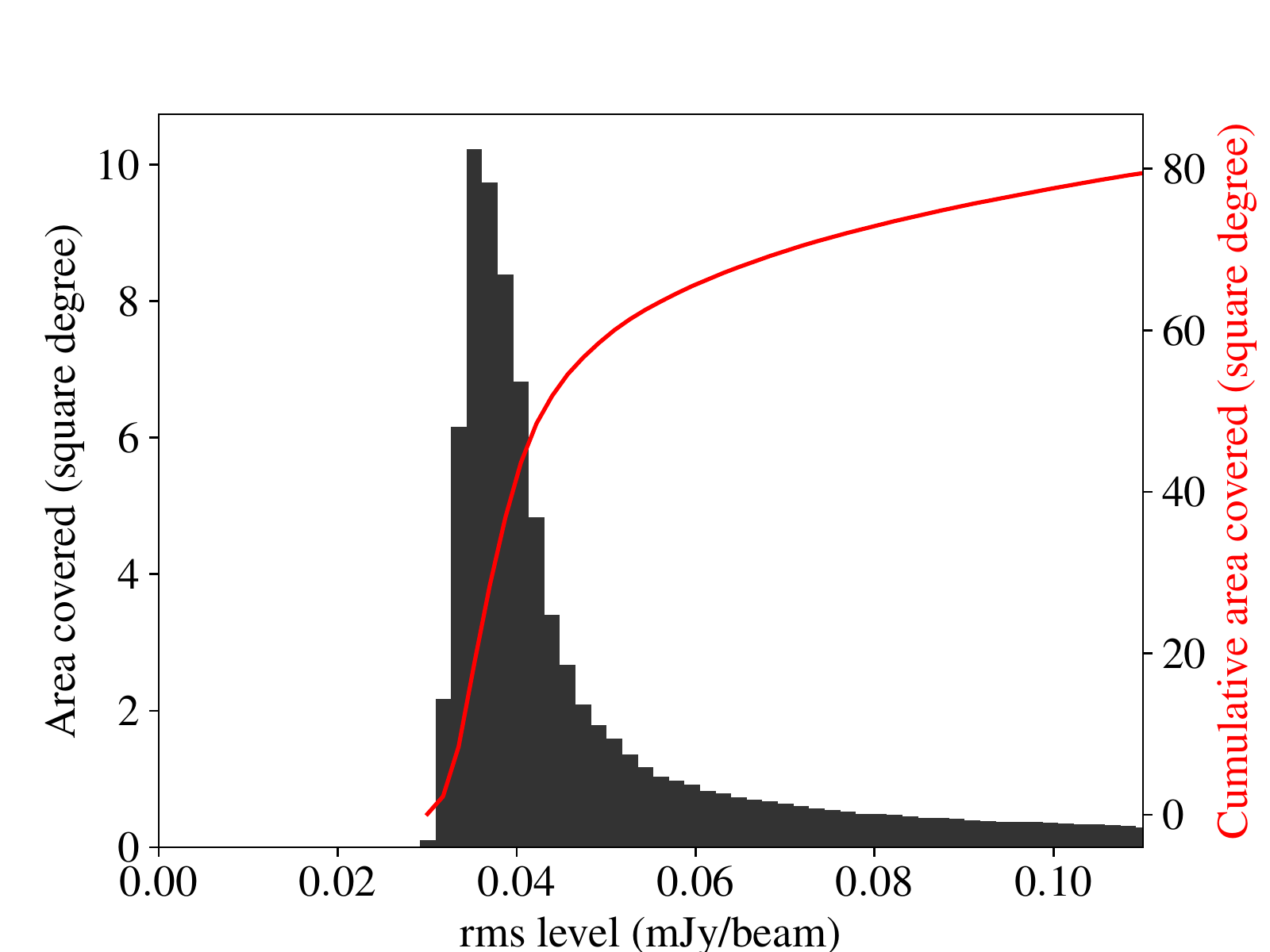}
    \caption{The histogram shows the distribution of the noise values in the noise map. The red line shows the cumulative area of the map with noise less than a given value.}
    \label{fig:visf}
\end{figure}

ASKAP data of the G23 field were reduced using $\texttt{ASKAPSoft}$ (Version 0.24.0) in 2019. Detailed information about the ASKAP data processing can be found in the $\texttt{ASKAPSoft}$ web page \footnote{\href{https://www.atnf.csiro.au/computing/software/askapsoft/sdp/docs/current/index.html}{https://www.atnf.csiro.au/computing/software/askapsoft}} (see also \citealt{mcconnell20}). Here we briefly describe the steps taken in processing the data. As a first step, anomalous samples in the data from both calibrator and science fields were identified and removed. Frequency dependent gains have been estimated using a primary flux calibrator (PKS B1934$-$638) for each beam, which were then used to set the instrumental flux-density scale and interferometer phase for each beam centre. Gridding and imaging of the calibrated multi-frequency data for each beam (using a robust parameter of -0.5) has been done using $\texttt{cimager}$ (version 1.0.1) in parallel processing on the Pawsey \footnote{\url{https://pawsey.org.au/}} computing cluster. A field model obtained for each beam as part of the imaging step has been used to perform self-calibration (this has been repeated twice). Images for each beam were made with a size of 4096\,x\,4096 pixels (each pixel is 2.5 arcsec). Standard continuum imaging parameters along with a Gaussian taper preconditioning (which allows to fit the beam while taking into account the restoring beam size defined by the user) were used to produce final images for each beam. The 36 final images were then primary beam corrected and mosaiced together in a single tile and the two resulting tile mosaics were further mosaiced together using the ASKAPSoft mosaicing function to generate the full radio map of the G23 field. The mosaicing function assumes a circular beam of full-width half-maximum (FWHM) 1.09$\lambda$/D for the primary beam model of the individual 36 beams. In addition, while combining individual beams, pixels whose primary beam gain is less than 20 per cent were disregarded.  The final mosaic image has a 10 arcsec resolution and a central rms level of 38$\mu$Jy beam$^{-1}$ which is relatively smooth across the image. In Fig. \ref{fig:rms} we show the rms noise map of the full mosaic and in Fig. \ref{fig:visf}  we show a histogram of noise estimates within the mosaic image. The red line shows the cumulative area of the mosaic with a noise less than a given value. 

\section{Data quality}\label{sec:3}

\subsection{Source extraction and flux scale assessment}\label{extract}
The radio sources were extracted using the Python Blob Detector and Source Finder (PyBDSF, \citealt{sf}). PyBDSF was run on the final mosaic using the parameters given in Table \ref{tab:psparams}. The good surface brightness sensitivity of ASKAP allows us to observe several extended sources with a variety of morphologies. We utilise the \texttt{trous} wavelet module of PyBDSF which is designed to recover the low surface brightness extended emission as much as possible. Briefly, the residual image which results from the fitting of Gaussian components is decomposed into wavelet images of various scales. This decomposition by the wavelet module aims to recover any extended emission that is not well fit during Gaussian fitting. In practise, any remaining emission in the residual image can be further fit by Gaussian functions whose size is tuned to the various wavelet scales. The PyBDSF parameters used for the source detection in our work were defined based on earlier works published in literature \citep[e.g.][]{ww19}, and were fine-tuned by running some tests and evaluating the results of these tests. PyBDSF produces a catalogue of individual Gaussian components and a catalogue of sources. The latter is formed by grouping Gaussian components together, when they belong to the same 'island'. An island is defined as a region of adjacent pixels, where the flux is above a certain threshold. This allows PyBDSF to successfully combine components associated to the same source, as far as such components are nearby or overlapping. If source components are well separated, a visual inspection is necessary to identify them and merge them into a single multi-component radio source (see Section \ref{sec-finalcat}). Both catalogues provide a number of source/component parameters. These include the position, integrated and peak flux density, structural parameters (measured and deconvolved sizes) and their estimated errors. The initial (raw) PyBDSF source catalogue consists of 54814 sources with 40186 objects having a peak flux density above a threshold of 5$\sigma$. 

\begin{figure*}
	\includegraphics[width=180mm]{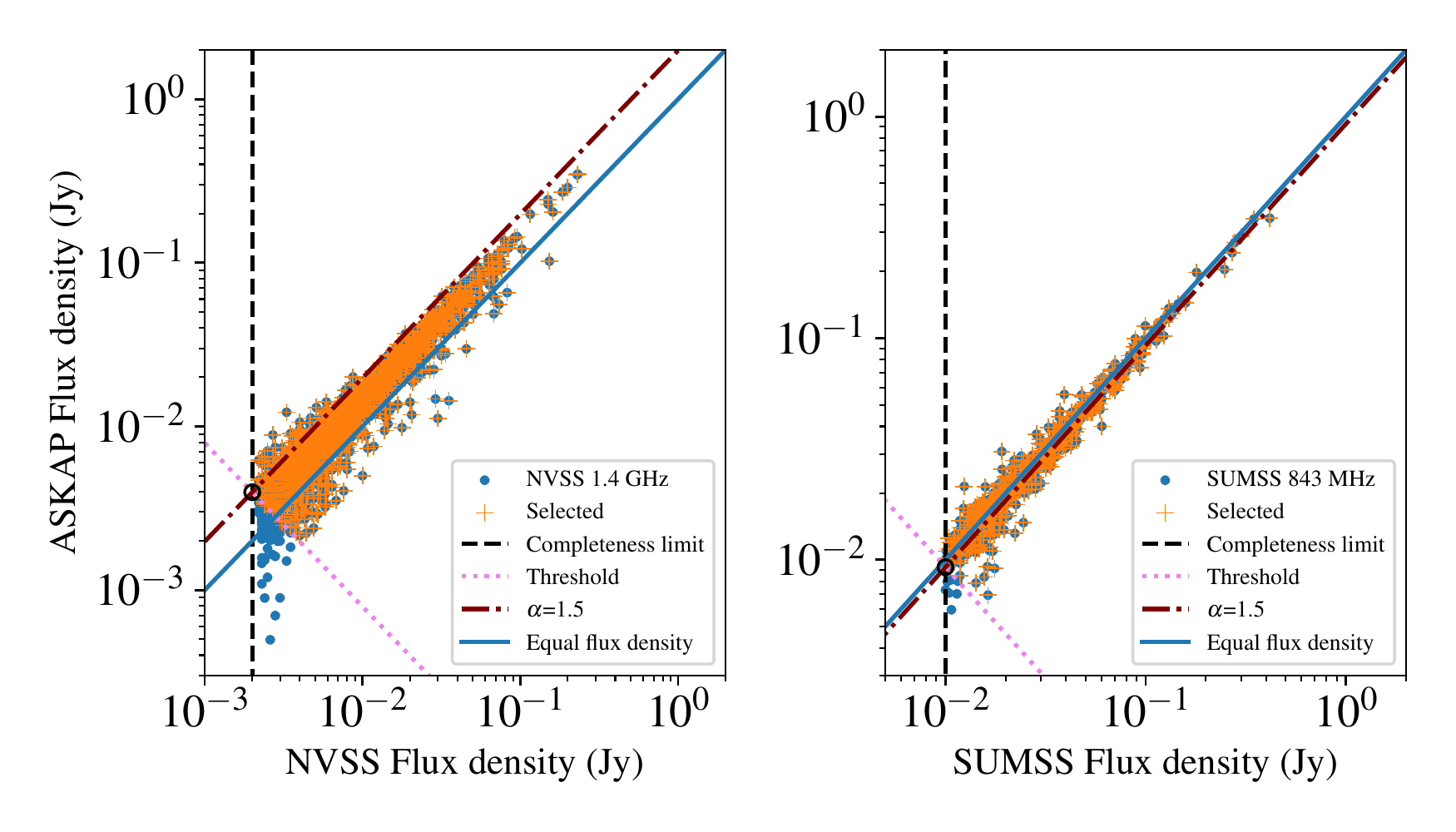}
    \caption{Selection of sources for the cross-match for the NVSS 1.4 GHz (left panel) and the SUMSS 843 MHz (right panel) samples.  All the cross-matched sources are shown as blue dots and the selected ones after various filtering are marked with an orange cross. The survey's completeness limit is shown as a vertical dashed line. The line at which the flux densities are equal is shown as a solid blue line. The locus of sources with a spectral index of 1.5 is shown as a dash-dotted maroon line. The point at which this line and the survey completeness line cross is used as a reference for the selection threshold in the product of flux densities (see text) which is marked as a pink dotted line.}
    \label{fig:fluxrat}
\end{figure*}

In order to evaluate the accuracy of the mosaic flux scale, external radio catalogues from various surveys covering the G23 field were utilised. We particularly exploited all the available surveys, independent of their observing frequency. In this way, we avoid relying our comparison on a single survey and on a specific spectral index assumption. Following \citet[][]{js21}, we compared the flux densities of sources in common between our catalogue and catalogues from  reference surveys, like: SUMSS \citep{sumss03}, NVSS \citep{nvss98}, TGSS \citep{tgss17}, GLEAM \citep{gleam17} and RACS \citep{mcconnell20,hale21}. Cross-matching of our source catalogue with external catalogues has been performed using different matching distances based on the spatial resolution of the comparison surveys. A number of constraints have been applied to external catalogues in order to eliminate possible biases that might be caused by the different effective depths and angular resolutions of the surveys. Parameters used for filtering and cross-matching are shown in Table \ref{tab:fsparams}. We initially selected isolated ASKAP sources (with a minimum distance to the nearest ASKAP source of 60 arcsec) in order to avoid contamination from neighbour sources. To avoid effects caused by different resolutions the matching was limited to only compact ASKAP sources (i.e. size of deconvolved major axis$<$15 arcsec). In order to minimize incompleteness effects only sources down to the completeness flux density limit of each comparison survey were taken into account. An additional survey dependent minimum flux density threshold was applied. A further constraint was applied in the product of flux densities of the ASKAP measurement and each comparison survey because applying a flux density cut in only one of the reference surveys may introduce a bias towards sources with high absolute values of their spectral indices. We defined the threshold by multiplying the flux density of a source at the completeness limit of the comparison survey by the ASKAP flux density that such a source would have a spectral index of 1.5 at the ASKAP observing frequency (887.5 MHz). The selected thresholds and completeness limits for all comparison surveys are listed in Table \ref{tab:fsparams} and the illustration for two surveys are shown in Figure \ref{fig:fluxrat}. The maximum cross-match distance to the comparison survey was empirically determined based on the resolution of the matched survey (for lower resolution surveys larger offset values were used). 

\begin{figure}
	\hspace{-0.6em}\includegraphics[width=98mm]{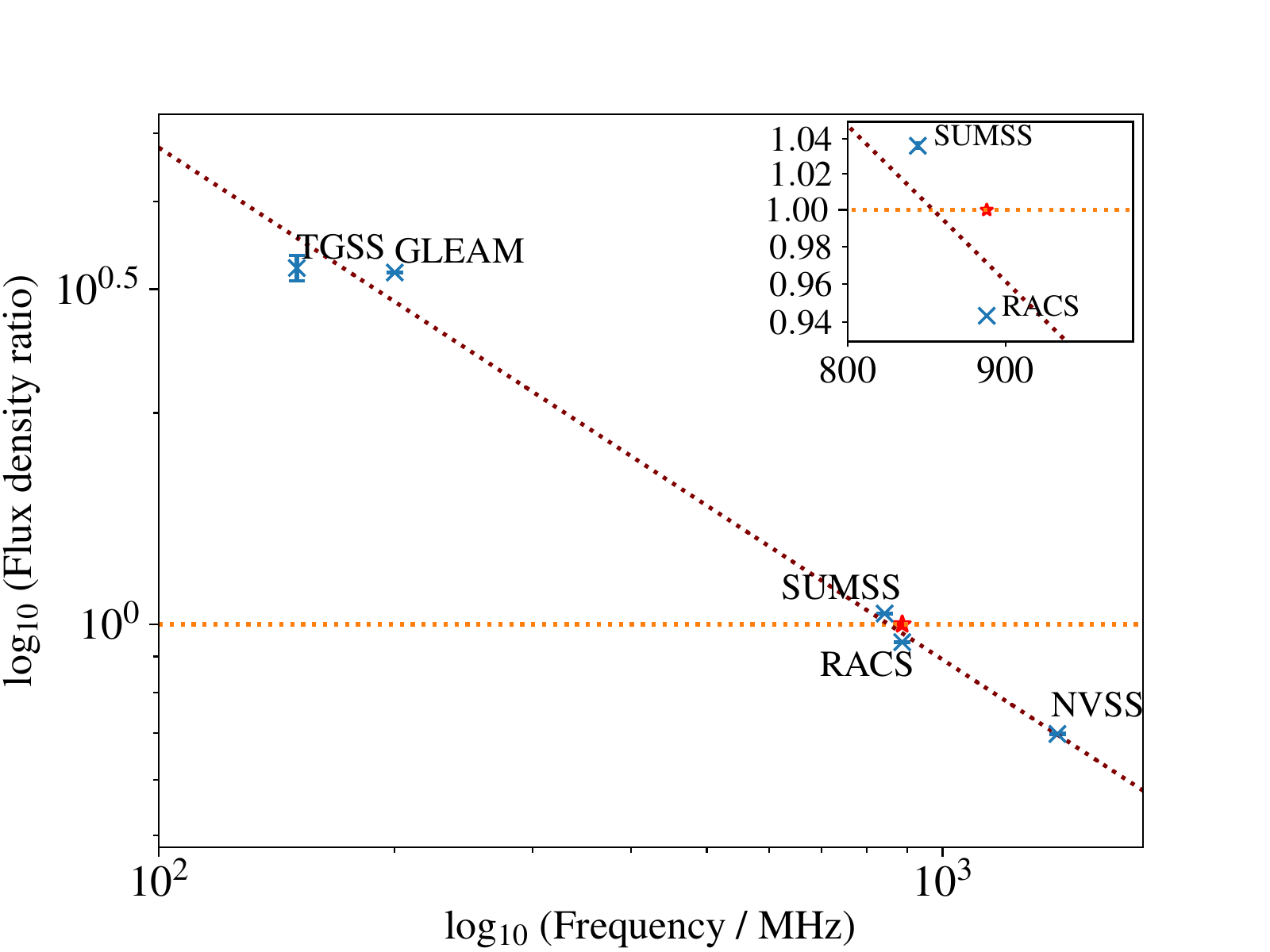}
    \caption{The median ratio of ASKAP total flux density to other survey total flux density are shown as a function of frequency of the surveys used for the cross-matching. The red star indicates the current ASKAP flux scale and the orange dotted line shows a ratio of unity whereas the maroon dotted line shows the best fit to the data points. The inner panel shows a zoomed version of the plot in the 800$-$1000 MHz range.}
    \label{fig:fluxrat2}
\end{figure}

Using the filtered sources we estimated the median values of the ratios of ASKAP total flux densities to other survey flux densities and these are shown in Fig \ref{fig:fluxrat2}. Errors in the median ratios were estimated using a bootstrapping technique. It is known that TGSS has region dependent issues with the flux density scale (e.g. \citealt{tgss17}). Therefore, we took into account the flux density uncertainty associated to a survey by adding it in quadrature to the error of the median of the flux density ratio. We used an Orthogonal Distance Regression \citep{isobe90,br90} fit that takes into account the uncertainties in the ratios to estimate the flux density scale ratio as a function of frequency. Using the values obtained from the fitting process we obtained a ratio of 1.016 ($\pm$0.11) at 887.5 MHz. As can also be seen in Fig. \ref{fig:fluxrat2}, our current flux density scale is very close to the best fitting line (the dotted maroon line in Fig. \ref{fig:fluxrat2}). It is reassuring that overall our flux densities agree well with those of all the other surveys, including those undertaken at nearby frequencies, such as RACS and SUMSS. In particular SUMSS fluxes are in perfect agreement with ours, while the discrepancy with RACS is around 5 per cent. This level of discrepancy is fully acceptable, given that the RACS survey shows position-dependent flux scale offsets up to 20 per cent, when compared with SUMSS \citep[][see their Fig. 10]{hale21}. Since the offset from the best fit line is only 2 per cent we do not further scale the flux density of our G23 sources.

\begin{table}
	\centering
	\caption{Key PyBDSF parameters that were used to obtain a catalogue of radio sources. $\texttt{adaptive thresh}$ sets the signal-to-noise (SNR) above which sources may be affected by strong artefacts, $\texttt{atrous jmax}$ gives the maximum order of the atrous wavelet decomposition, $\texttt{group tol}$ is the tolerance parameter for grouping of Gaussians into sources, $\texttt{rms box}$ sets the box size and step size for rms/mean map calculation, $\texttt{rms box bright}$ sets the box size and step size for rms/mean map calculation near bright sources and $\texttt{thresh isl}$ determines the extent of island used for fitting.}
	\label{tab:psparams}
	\begin{tabular}{lc}
		\hline
		\hline
        $\texttt{adaptive thresh}$ & 150\\
        $\texttt{atrous jmax}$ & 4\\
        $\texttt{group tol}$ & 10\\
        $\texttt{rms box}$ & (150, 15)\\
        $\texttt{rms box bright}$ & (60, 15)\\
        $\texttt{thresh isl}$ & 4\\
		\hline
	\end{tabular}
\end{table}

\begin{table*}
	\centering
	\caption{Parameters used for the cross-match with external surveys and their filtering. The columns are respectively: (1) Survey name. (2) Reference survey frequency. (3) Resolution of the reference survey. (4) Maximum cross-match distance for the survey. (5) Survey completeness limit. (6) Maximum size of the deconvolved major axis used for filtering. (7) Value of the selection threshold in the product of the flux densities (see text). (8) Number of sources selected after the cross-match and filtering.}
	\label{tab:fsparams}
	\begin{tabular}{lccccccc} % four columns, alignment for each
		\hline
		\hline
		Survey & Survey&Survey & Max. cross-match & Survey flux & Max. major & Flux product & N\\
		       &Frequency& resolution & radius & density limit  & axis size   & threshold & \\
		       && (arcsec) & (arcsecs) & (mJy) & (arcsecs)  & (mJy)\\
		\hline
		NVSS&  1.4 GHz & 45 & 7  & 2  & 15 & 3.963 & 1410\\
		SUMSS& 843 MHz & 45 & 7  & 10 & 15 & 9.257 & 454\\
		TGSS&  150 MHz & 25 & 6  & 30 & 15 & 2.084 & 545\\
		RACS&  887.5 MHz & 25 & 6  & 1  & 15 & 1.000 & 226\\ 
		GLEAM& 200 MHz & 130 & 20 & 50 & 20 & 5.349 & 486\\
		\hline
	\end{tabular}
\end{table*}

\subsection{Astrometric precision}
In order to evaluate the astrometric precision of ASKAP source positions we cross-matched the initial PyBDSF catalogue with various radio survey catalogues and the GAMA optical catalogue and calculated the positional offsets. These offsets are reported in Table \ref{tab:poparams}. The astrometry of GAMA is tied to the KiDS survey in the G23 field. KiDS has internal astrometric error $<$0.03" \citep{kids13} and more robust than any other reference radio survey used in this work. The median ASKAP-GAMA position offsets are around 0.2 arcsec and 0.1 arcsec for RA and DEC, respectively (see Table \ref{tab:poparams}). We applied these offsets to the source positions in order to bring our sources to a better agreement with the GAMA survey.  While PyBDSF estimates uncertainties in source positions using the method given by \citet{condon97} these still can be underestimated \citep[e.g.][]{condon15}. To further evaluate uncertainties in source positions we select bright and compact radio sources and estimate the angular separation from their hosts (see Section \ref{sec-finalcat}), in both RA and Dec. The error in the mean separation is around 0.6 arcsec for both RA and Dec and this may be a better representation of the uncertainty in radio source positions.
%The offsets between ASKAP detected sources and their GAMA counterparts in Fig. \ref{fig:optpos}, and provide  Given the resolution of the ASKAP G23 survey which is 10 arcsec such offsets are insignificant and can be neglected.

\begin{table*}
	\centering
	\caption{Median astrometric offset from other reference surveys and the bootstrap errors of these offsets for RA and DEC, respectively.}
	\label{tab:poparams}
	\begin{tabular}{lccccc} % four columns, alignment for each
		\hline
		\hline
        Surveys & Survey& RA & DEC & Error on median RA & Error on median DEC\\
		&resolution & offset & offset &&\\
		&(arcsec) &(arcsec)  & (arcsec)  &&\\
		\hline
        GAMA  & 2 & -0.18& 0.1& 0.008 & 0.006\\
        NVSS  & 45 & -0.48& -0.11& 0.04 &0.05\\
        SUMSS & 45 & 0.33& -0.16& 0.11 & 0.1\\
        RACS  & 25 & -0.005& -0.008& 0.003 & 0.003\\
        TGSS  & 25 & -0.29& -0.28& 0.04 & 0.06\\
        GLEAM & 130 & -0.90& -3.04& 0.27 & 0.23\\
		\hline
	\end{tabular}
\end{table*}

\subsection{Completeness, reliability and bandwidth smearing}\label{sec:compl}
EMU Early Science observations of the G23 field were carried out with a frequency resolution of 1 MHz, so some level of bandwidth smearing is expected, causing an underestimation of the source peak flux density and a corresponding overestimation of source sizes in the radial direction such that the integrated flux density remains unchanged. Bandwidth smearing increases with the angular distance from the pointing centre of phase and depends on the bandpass width, the observing frequency, and the synthesised beam size. In a single ASKAP beam the peak flux reduction due to bandwidth smearing is expected to increase radially to around 8 per cent at $\sim$ 1.7 deg angular distance from the phase centre (i.e. at a distance equal to the FWHM of the primary beam). The observations were done with a beam spacing of 0.9 degrees. Such a spacing yields a uniform sensitivity across the observed region, when all beams are mosaiced together in a tile. The mosaicing consists of a weighted linear combination of the pixels of the overlapping beams, where the weights are estimated by taking into account beam response curves and local noise. As extensively discussed by \citet[][]{prandoni2000}, such a combination results in a nearly uniform smearing across the mosaic (or tile). In other words the radial dependence tends to cancel out, due to the fact that pixels at different distance from their respective beam center are summed up. As demonstrated by \citet[][]{prandoni2000}, the expected smearing in the mosaic should be of the order of the one expected for sources located at maximum distance from the closest beam center, corresponding to $0.9\cdot \sqrt{2}/2 \sim 0.64$ degrees in our case (see \citealt{prandoni2000} for more details). We then expect a $\sim 2$ per cent reduction in the peak flux density due to smearing across the observed region.

\begin{figure*}
	\hspace*{-1em}\includegraphics[width=32em]{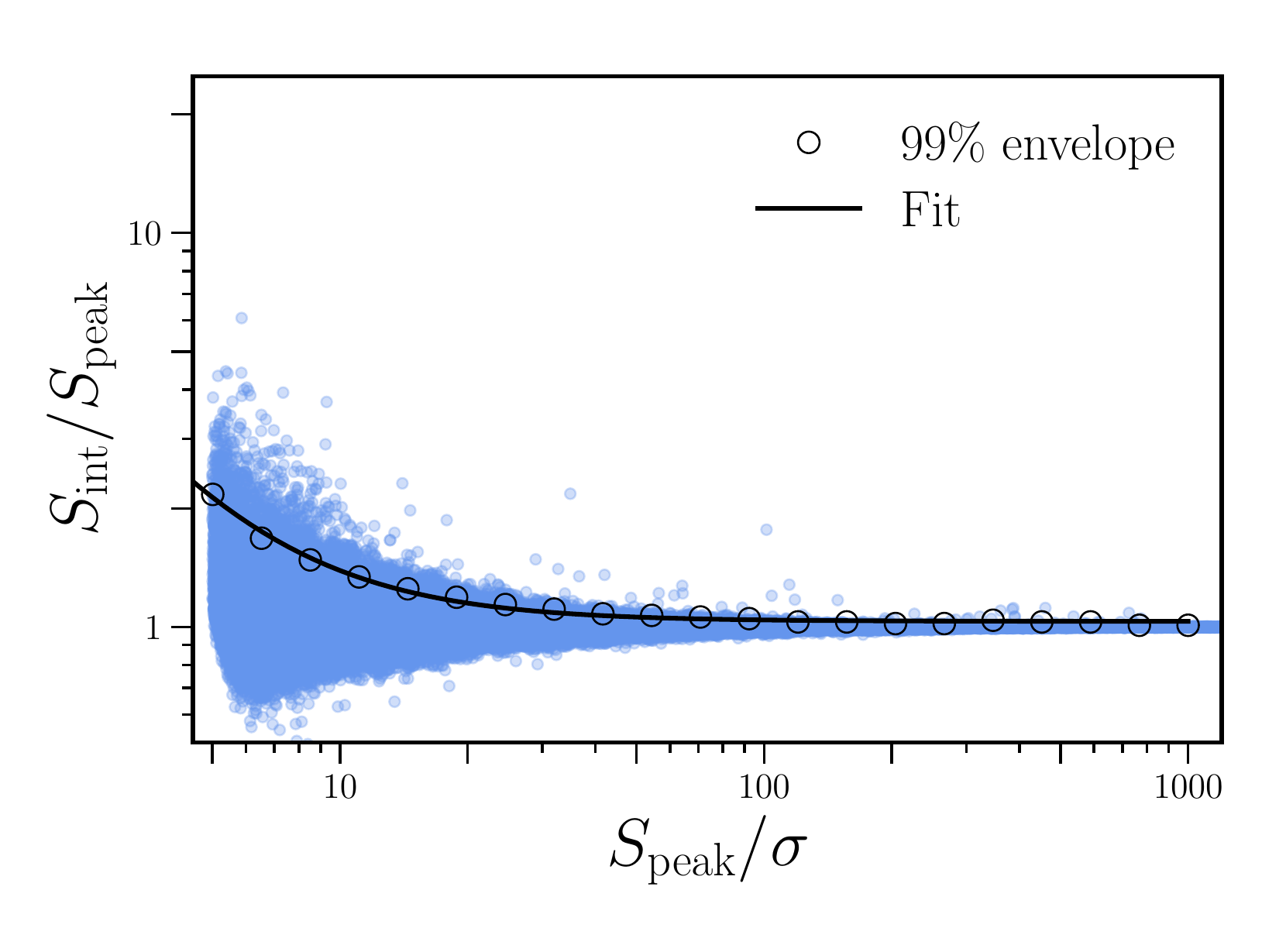}
	\includegraphics[width=31em]{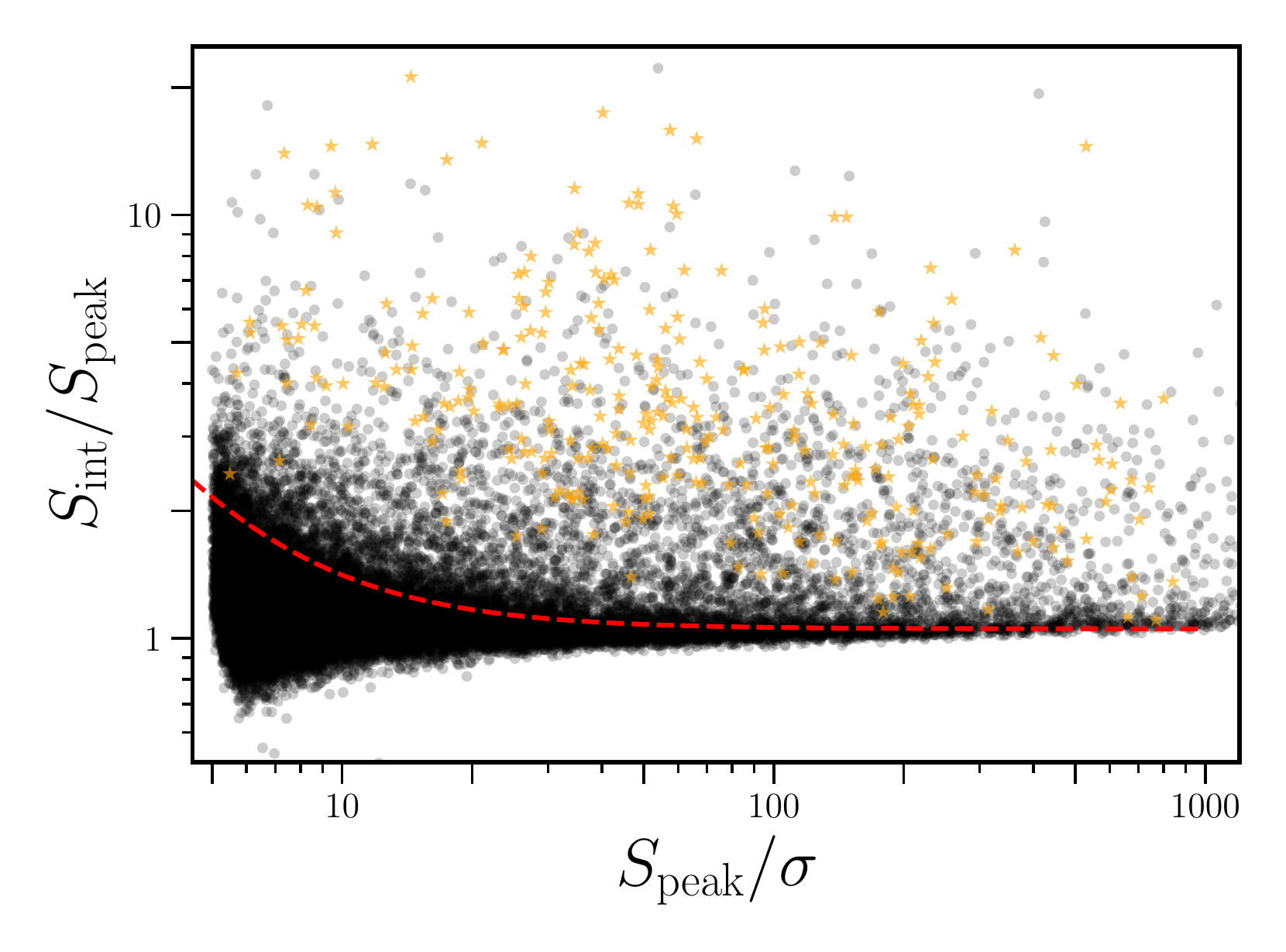}\\
    \caption{Left: Flux density ratio of the total flux to peak flux of simulated sources as a function of SNR. The black points show the threshold below which 99 per cent of the sources lie in 20 SNR bins. The black line shows a fit to this upper envelope. Right: Flux density ratio of the total flux to peak flux of sources in the source catalogue as a function of SNR. The red dashed line shows the modified version of the fitted envelope (see text) and orange stars are multi-component sources which are identified via visual inspection. Below the red line sources are defined to be unresolved and as expected all multi-component (i.e. resolved) sources are above this line.}
    \label{fig:sim1}
\end{figure*}

In order to better characterise the sources and evaluate completeness of the survey, we carried out Monte Carlo simulations in the image plane. Point sources at random positions were injected in the residual map produced by PyBDSF (this ensures that the noise and its spatial distribution are consistent with that of the real data), and were recovered with PyBDSF using the same set of parameters as used for the real cataloguing. We chose the injected source flux densities randomly, using a power law slope, spanning the range of observed flux densities in the survey (starting from 0.0001 Jy). We carried out 10 set of simulations for each of a range of input source flux densities, using around 24,000 sources per run to improve the statistics. On the left panel of Fig. \ref{fig:sim1} we show the flux density ratio of total flux to peak flux of recovered sources as function of signal-to-noise ratio (SNR). For point sources the ratio of total to peak flux density is expected to be unity and the observed spread around unity is due to measurements errors. We determined the 99 per cent envelope by fitting the 99th percentile in 20 logarithmic bins across the range of SNRs (seen as black open circles in the left panel) using the simulated sources. The fitted envelope is described by the function 1.03$\pm$0.007+15.01$\pm$1.77/SNR$^{1.62\pm 0.07}$ (black solid line in the left panel of Fig. \ref{fig:sim1}). It is clear that the total-to-peak flux density ratio of the simulated point sources tends to a value higher than unity ($\sim$1.03) for high SNRs. Our investigation showed that this is due to the fact that the detection algorithm (PyBDSF) tends to overestimate the sizes of point sources. By repeating this analysis on the real source catalogue we find that the total-to-peak flux density ratio tends to a slightly higher value (1.05) for large SNRs, and is better represented by the modified envelope 1.05+15.01/SNR$^{1.62}$ (see red dashed line in the right panel of Fig. \ref{fig:sim1}). This extra 2 per cent offset is fully consistent with the expected bandwidth smearing, but we cannot exclude some contribution from possible residual phase errors due to the lack of implementation of direction-dependent calibration \citep[e.g.][]{wilber20}. We also note that we examined the total-to-peak flux density ratio of high signal-to-noise point-like sources in different regions of the mosaic, and found no systematic trends across the mosaic. We utilise the envelope function shown in the right panel of Fig. \ref{fig:sim1} to identify reliably resolved sources. All sources below that function are considered to be unresolved. Sources with multiple components (i.e. extended sources), obtained as part of the visual inspection (see Section \ref{sec-catalogue}), are shown as orange stars. All of them are above the envelope, as expected.

\begin{figure*}
	\includegraphics[width=31em]{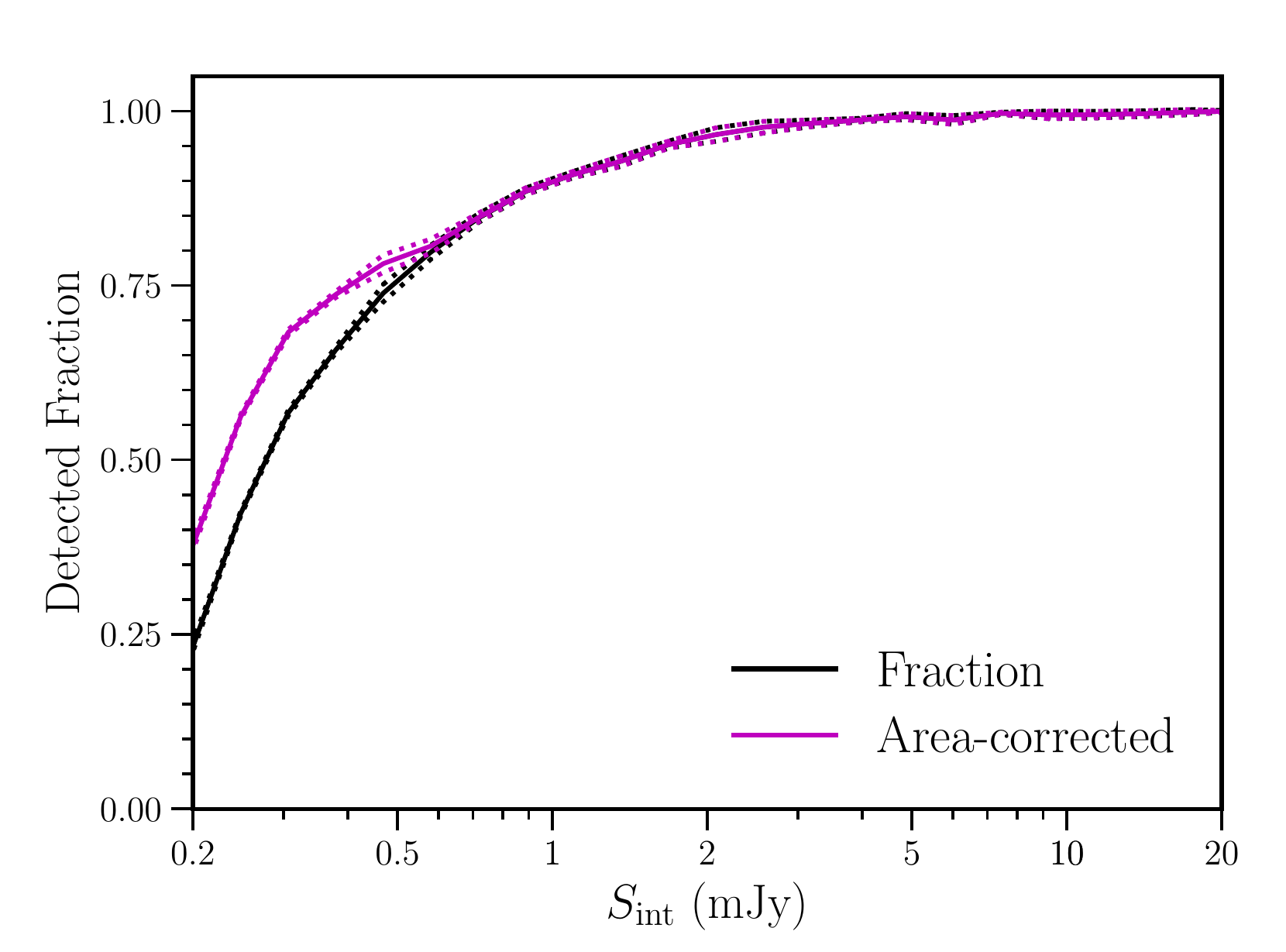}
	\includegraphics[width=31em]{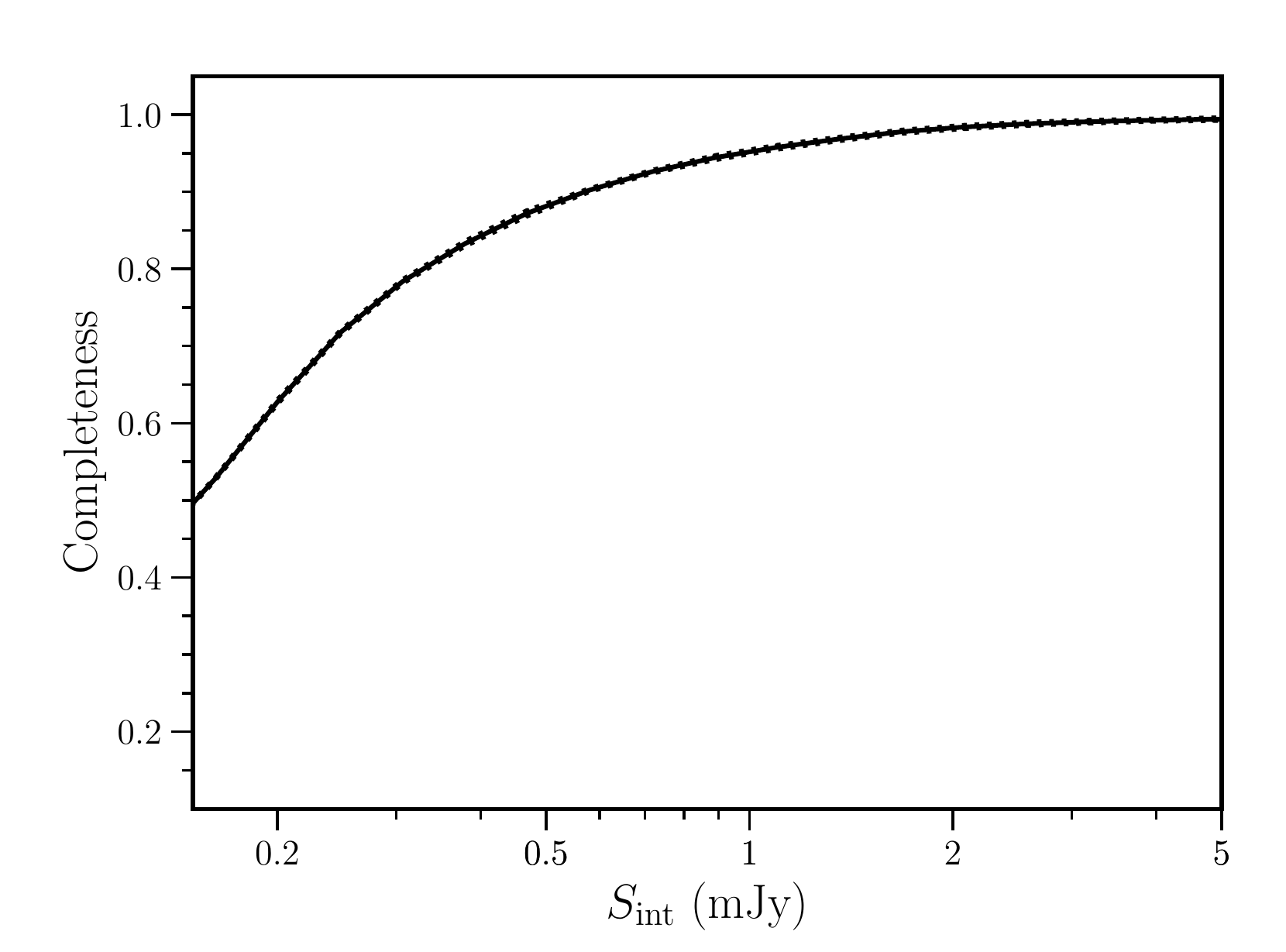}\\
    \caption{Results of the Monte Carlo completeness simulations. Left: Fraction of sources detected as a function of total flux density. The purple curve shows the detected fraction correcting for the effect of the visibility area. Right: Results of the completeness analysis.  The dotted lines in both panels show the 1-$\sigma$ uncertainties.}
    \label{fig:cness}
\end{figure*}

We have estimated the completeness of the catalogue (i.e. the probability that all sources above a given flux density are detected) by computing the fraction of detected sources in our simulation as a function of total flux density and integrating the detected fraction upwards from a given flux density limit. The detection fraction of sources as a function of total flux density is shown as a black line in the left panel of Fig. \ref{fig:cness}. This detection fraction is expected to be largely driven by the variation in rms across the image (visibility area$-$the fraction of the image over which a source of a given flux density should be detectable). The number of detected sources as a function of sources that could be detected, accounting for the visibility area, is also shown in the left panel as the purple line. The results of the completeness analysis (that takes into account the visibility function) are shown in the right panel of Fig. \ref{fig:cness}. The catalogue is expected to be 90 per cent complete above 0.5 mJy for point sources. The source peak flux densities are reduced due to smearing (while total flux densities are not affected by the smearing effect). This will introduce additional incompleteness in the catalogue, as source detection is based on signal-to-noise thresholds (SNR=S$_{\rm peak}/\sigma$). Considering that smearing is expected to be around 2 per cent, we expect extra incompleteness at SNR$< 5.1\sigma$).

\begin{figure}
	\includegraphics[width=90mm]{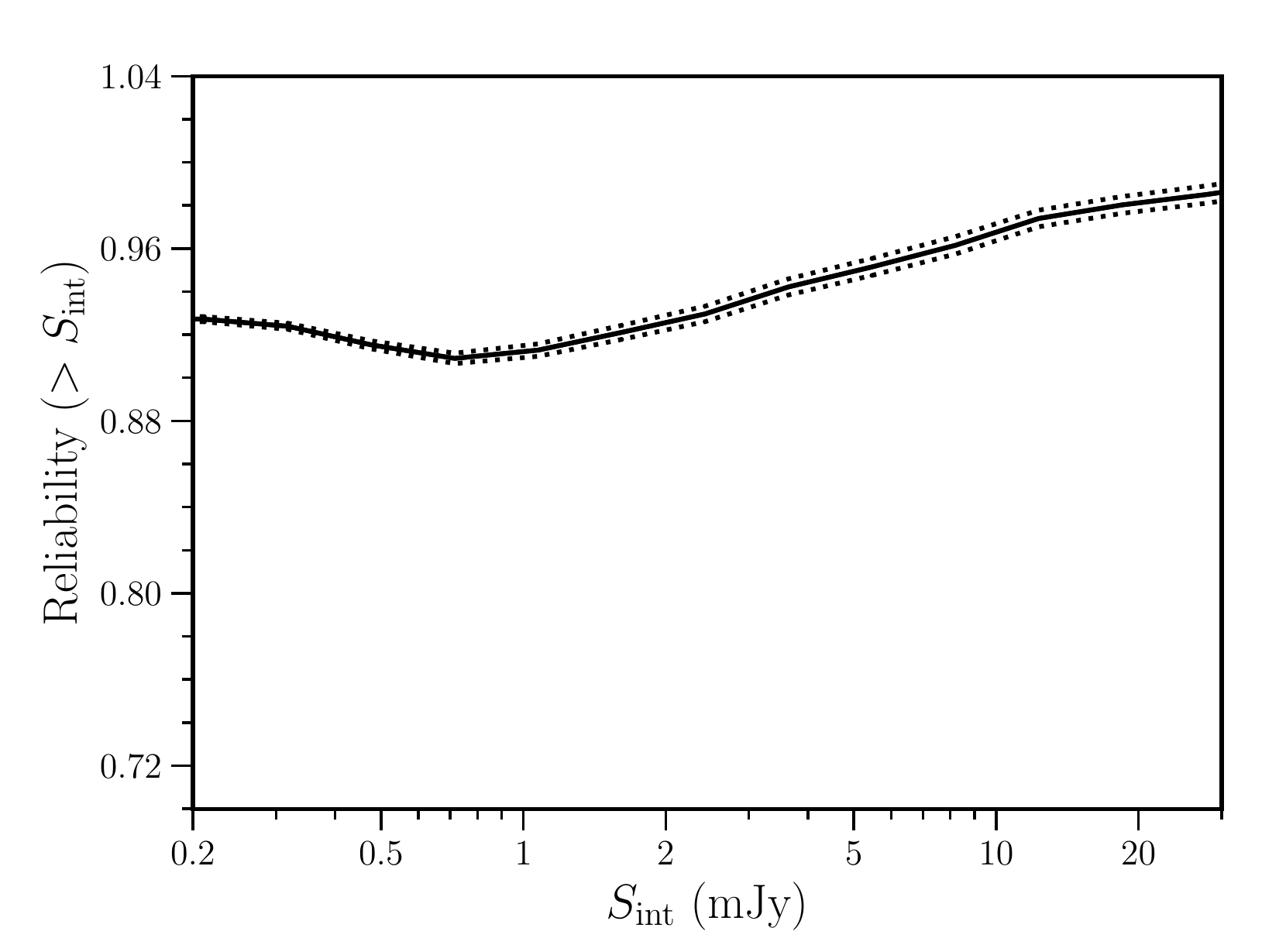}
    \caption{Results of the estimated reliability of the catalogue as a function of integrated flux density limit. Here the reliability is defined as the probability that all sources above a given total flux density are real. The reliability is $>90\%$  over the entire flux density range. The dotted lines show the 1-$\sigma$ uncertainties.}
    \label{fig:rely}
\end{figure}

The reliability of a source catalogue can be defined as the probability that all sources above a given total flux density are real. In order to estimate the reliability, we extracted sources from the inverted residual mosaic image (assuming that negative image background features are statistically the same as positive ones). The detected negative ‘sources’ were grouped by total flux density into 20 logarithmic bins and these were compared to the binned results of the positive (i.e. detected in the mosaic by a regular source extraction process; see Section \ref{extract}) sources. For our positive sources we use the initial catalogue. The real number of sources are defined to be the number of positive sources after extracting the number of negative sources. This analysis showed that the reliability is $>$90 per cent down to that faintest flux densities. Fig. \ref{fig:rely} shows the reliability curve determined from the number ratio of real sources over positive sources above a given total flux density. 

It is worth noting that the completeness and reliability analyses were carried out for point sources only because at the faint end of the catalogue, where the effects of completeness and reliability are important, a large fraction of the sources are point-like. Taking into account the modified envelope (see Fig. ~\ref{fig:sim1}, right panel) 70 per cent are point-like below 1 mJy (i.e. at the fluxes where we see incompleteness). We expect spurious sources associated with  noise peak detections to be point-like and dominate close to the detection threshold. Artefacts around bright sources may not be point like and would contribute to the curve given in Fig. \ref{fig:rely} at fluxes larger than 1 mJy (and they were indeed removed, see Section \ref{sec-finalcat}).

\section{Final data products and sample properties}\label{sec:4}
\subsection{Multi-wavelength data and catalogues}\label{sec-catalogue}

This paper uses a multi-wavelength catalogue to identify the host galaxies of radio sources as described in Section \ref{sec-finalcat}. The multi-wavelength catalogue has been created by combining together different data products made available by the GAMA collaboration \citep[]{driver11} that granted us advanced access to the GAMA DR4 catalogues (Driver et al. submitted). To construct our multi-wavelength catalogue we combined the following GAMA catalogues: the photometric catalog \texttt{KidsVikingGAMAV01} produced with the \texttt{ProFound} image analysis package \citep{robotham18} described by \cite{bellstedt20}, the spectroscopic redshifts catalogue \texttt{SpecObjv27} \citep[]{liske15} and the spectroscopic line measurements from the \texttt{GaussFitSimplev05} and \texttt{GaussFitComplexv05} catalogues \citep[]{gordon17}. The combination of these GAMA products provided a catalogue with (i) GALEX- FUV and NUV, (ii) VST KiDS $u,g,r,i$ (iii) AllWISE- 3.4, 4.6, 12, 22$\mu$m, (iv) $Herschel$ SPIRE 250 350, 500 $\mu$m photometric bands information and spectroscopic information with a 95 per cent completeness down to the VST KiDS $i$ magnitude of 19.2 \citep[]{liske15}, whereas this limit is i$<22.25$ for the photometric sample. 

The aforementioned catalogue was used to define a reliable sample of galaxies and AGN to be used for our source identifications and provided all the information needed to characterise radio sources with host galaxy counterparts as described in Section \ref{sec:sprop}. In order to remove the spurious sources and artefacts we followed the instructions provided by \cite{bellstedt20}. Since some radio galaxies may appear as star-like object in the optical we did not remove stars from the GAMA catalogue but we flagged these objects as star-like or galaxy-like using two classification methods given by \cite{baldry12} and \cite{bourne16}. This information assisted us in eliminating radio sources with host galaxies that are flagged as stars.

We also use photometric redshifts where spectroscopic redshifts are not available. To this end we exploit two photometric redshift catalogues: a photometric redshift (photo-z) catalogue provided by the  HELP collaboration \citep{shirley21}, (ii) a catalogue of photometric redshifts estimated as part of our collaboration (\texttt{gkvEAZYPhotozv02}, joined with the rest of the GAMA DR4 data products). We prioritise using the HELP catalogue for the reasons outlined below. The HELP catalogue provides reliable photo-zs for the sources that lack spectroscopic information. The HELP photo-zs are estimated using a public photometric redshift code EAZY \citep{eazy08} and following the methods described by \cite{duncan18a,duncan18b}. Briefly, the EAZY algorithm proceeds through a grid of redshifts defined by a user, and at each redshift it finds the best fitting synthetic template spectral energy distributions (SEDs) by minimising the $\chi^{2}$ goodness-of-fit parameter for every valid combination. The posterior probability density function (PDF) for redshift is then derived by considering the likelihood associated with the best fit template at each redshift, optionally with magnitude-dependent prior. The HELP collaboration uses different template sets: (i) stellar-only templates, (ii) the EAZY default library, two other sets including both stellar and AGN/quasar contributions and the SED templates by \citep{brown14}. In the estimation of photometric redshifts the HELP collaboration utilises either SED templates or machine-learning estimates. This produces a hybrid consensus photo-z estimate with accurately calibrated uncertainties. They also treat separately the known AGN samples (identified using optical, infrared and X-ray data) to obtain better estimates of AGN photometric redshifts.  

For the sources whose redshifts are not available in either catalogues mentioned above we then make use of a photometric redshift catalogue produced as part of our collaboration using the EAZY software. For this we used the atlas of 129 empirical galaxy spectra by \cite{brown14} as a template set. Since we do not treat AGN separately, photometric redshifts for sources where AGN emission dominates the observed optical$-$NIR spectrum might not be completely reliable. For a more comprehensive description of the multi-wavelength catalogues and multi-wavelength analysis of the radio sources we refer the reader to Marchetti et al. (in prep). 

\subsection{Source association, host galaxy identification and final catalogue}\label{sec-finalcat}

The initial radio source catalogue was processed in order to clean it from spurious sources and to reliably associate separate components belonging to the same source. To identify artefacts around bright sources we filtered the bright compact sources with multiple neighbouring sources, and visually inspected them. Confirmed artefact were then removed  from the catalogue. Then the source association and optical identification processes were carried out in parallel. As mentioned earlier for the host galaxy source identification we utilised the multi-wavelength catalogue described above. Since the multi-wavelength catalogue takes GAMA sources as its base there are GAMA sources with or without WISE counterparts. In the process of host galaxy identification if a radio source has a GAMA counterpart we take this as the host, and if not we take the WISE counterpart as the host.

Sources classified as 'single' (S$\_$Code=S) in the catalogue were initially cross-matched with the multi-wavelength catalogue assuming a matching distance of up to 2 arcsec. This value was chosen by estimating the incidence of false cross-identifications as a function of matching distance. We generated a mock catalogue of optical sources with random positions and cross-matched this mock catalogue with the radio source catalogue for a range of matching radii. We estimated the fraction of matches as a function of matching radius. Above a match radius of about 2 arcsec the number of matches stays approximately unchanged for any further increase in separation distance, and this is taken as the best pairing distance estimate. At 2 arcsec separation, the predicted contamination (N$_{\mathrm{random}}$/N$_{\mathrm{true}}$ where N$_{\mathrm{true}}$ is estimated by subtracting the total random matches from the corresponding value for the real catalogue) in the overall sample is around 7 per cent. Naturally, a larger matching radius would lead to a higher number of identifications and consequently to a more complete catalogue with identifications. At this adopted separation, the completeness rate (given by the fraction of true matches out of all matches) is around 85 per cent.

\begin{figure*}
	\includegraphics[width=60mm]{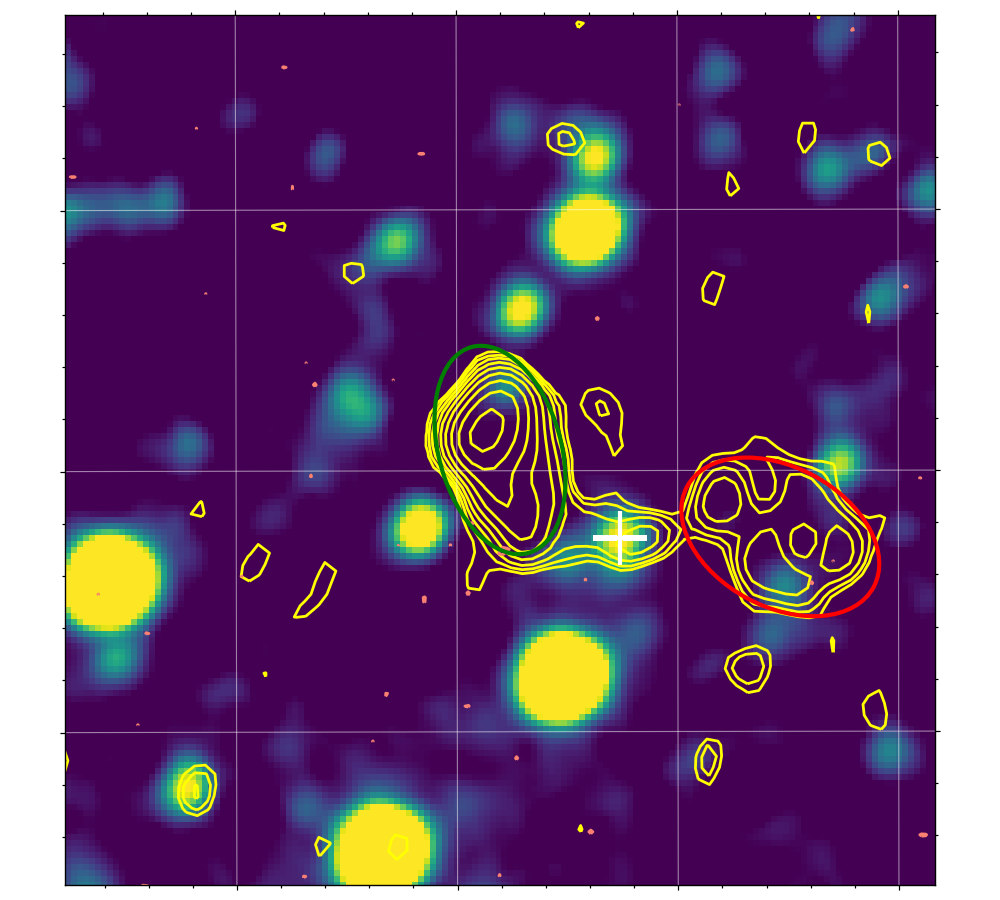}
	\includegraphics[width=60mm]{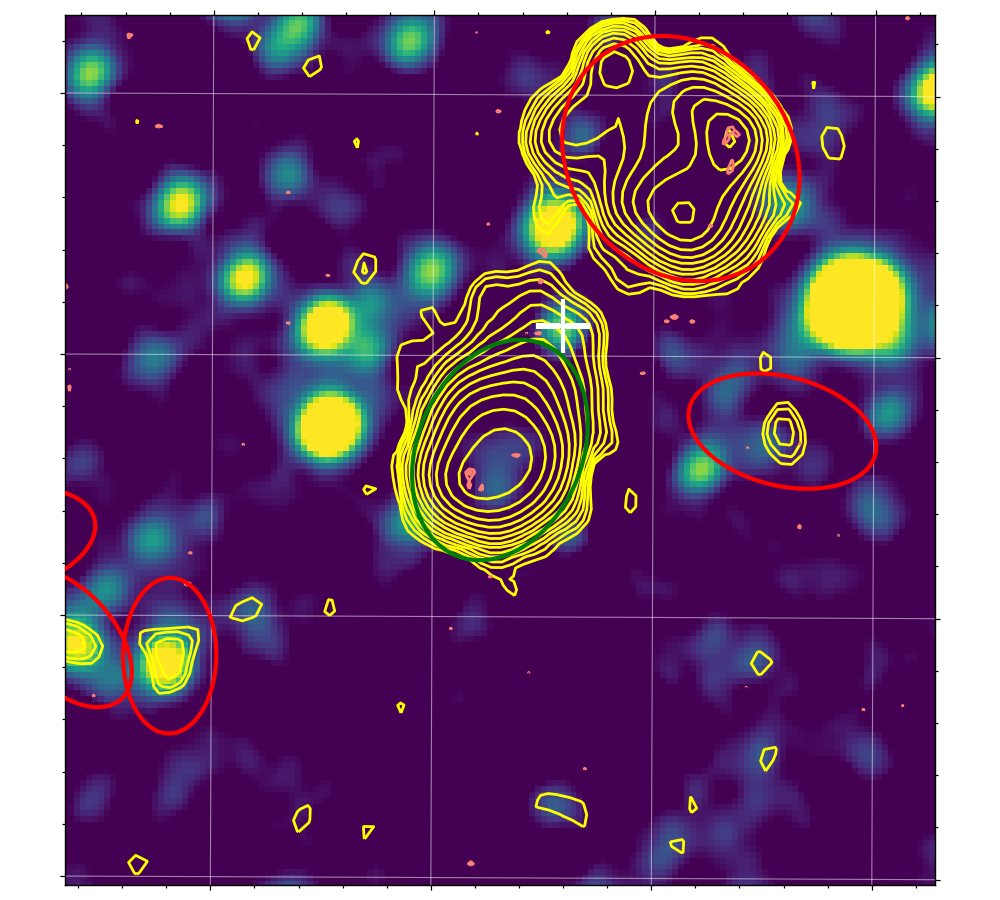}\\	
    \caption{Example postage stamps of radio sources in the G23 field. ASKAP contours (yellow) and VLASS contours (salmon) on WISE images (colour). Each postage stamp is centered at the position of the source component to be inspected (indicated by a green ellipse). The red ellipse shows the neighbouring source and the white cross indicates the host galaxy position. ASKAP and VLASS contours denote the surface brightness levels starting at 3$\sigma$ and increasing at various powers of 3$\sigma$ (off-source rms noise).}\label{postage}
\end{figure*}

The majority of the remaining sources were then visually inspected using the postage stamps (see \ref{postage}) that show ASKAP and VLASS contours on WISE images in order to identify and collect objects with multiple components (two examples of the postage stamps are shown in Figure \ref{postage}). The selected possible multi-component radio sources were run through another visual inspection utilising a tool, which displays the radio data as well as overlaying a catalogue of optical (GAMA) or infrared (WISE) sources around the target source to be classified. This process involved associating individually detected ASKAP sources, identifying any remaining artefacts, and, for real sources, specifying any plausible optical identification for the radio source. Here VLASS images had an important role, as they often show the flat-spectrum core of an extended ASKAP source making optical identification far more robust. 

We provide source sizes in the final catalogue along with source flux densities (in the case of multi-component sources a sum of component flux densities were stored), following a similar method presented by \citet{ww19}. We first determined the convex hull around the set of elliptical Gaussians: the convex hull is the smallest convex shape that contains all of the ellipses. We constructed the convex hull of multi-component sources by representing each component (PyBDSF source or PyBDSF Gaussian as appropriate) as an ellipse, where the deconvolved FWHM major and minor axes are taken to be, respectively, the semi-major and semi-minor axes of the ellipse. The convex hull was constructed around all of the component ellipses using the \textsc{shapely Python} package. The length of the largest diameter of the convex hull around the set of elliptical Gaussians (i.e. the maximum angular separation where all points on the convex hull considered pairwise) would give an estimate of a source size ("Length"). Sizes based on this presented method are likely to overestimate the true size in general. Another approach would be estimating the separation between the farthest components \citep[e.g.][]{hg16} but this will underestimate the true size (given as "Separation" column in the catalogue). For this reason we took the mean value of the length of the largest diameter of the convex hull around the set of elliptical Gaussians (i.e. Length) and the distance between the most separated components ("Separation") of a source as our true sizes ("Size"). The source position angle (given as "PA" column) was taken to be the position angle on the sky of that largest diameter vector. For the source width ("Width") we adopted twice the maximum perpendicular distance of points on the convex hull to the largest diameter vector. For the sources classified as 'single' we provided the major ("Maj"), minor ("Min") and deconvolved major ("DC$\_$Maj") and minor axis ("DC$\_$Min") of the Gaussian along with the errors on these measurements provided by PyBDSF, and its position angle. The error measurements do not take into account systematic effects (i.e. smearing which is expected to reduce peak fluxes by at most 2 per cent and ionosphere, see Section \ref{sec:compl}) which would affect the source size. This amount can be added in quadrature to the given errors. The estimation of the error in source size is less straightforward for extended radio sources (i.e. multi-component objects). The relative error on sources with deconvolved sizes larger than 20 arcsec (and detected at 10$\sigma$) is around 10 per cent and this is much lower for larger size objects. Based on this, we can safely suggest that sources with multiple components should have relative errors smaller than this value. It is worth noting that estimating both sizes and size errors is even more complicated for radio sources with bent morphology. The final catalogue has 39812 sources detected at 5$\sigma$ and 25087 (63 per cent based on the matching radius we use) objects have host galaxy counterparts (either GAMA or WISE). The number of unresolved sources, resolved sources and sources with multiple components are given in Table \ref{tbl:stat}.

\begin{table}
	\centering
	\caption{The source statistics based on the modified envelope (see Section \ref{sec:compl}) and the final catalogue.}
	\label{tbl:stat}
	\begin{tabular}{lc} % four columns, alignment for each
		\hline
		\hline
		Source type & Count \\
		\hline 
		Total unresolved sources &  32637\\
		Total resolved sources & 7175\\
        Visually identified sources with 1 component &175\\
        Visually identified sources with 2 components &86\\
        Visually identified sources with $\geq3$ components &44\\
		\hline
	\end{tabular}
\end{table}

\subsection{Source counts}

\begin{figure*}
	\includegraphics[width=190mm]{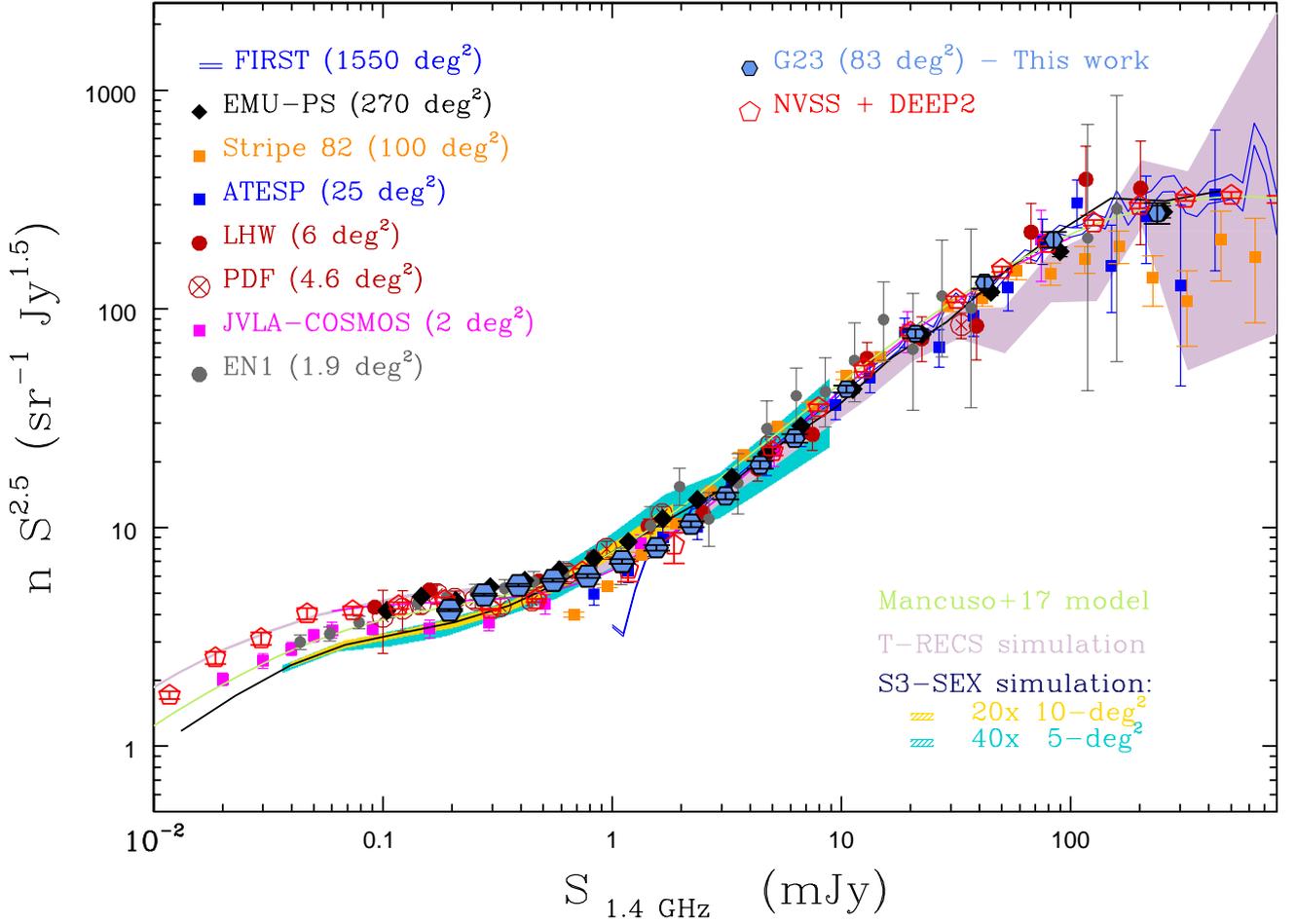}
    \caption{Normalized differential source counts derived from the G23 radio source catalogue (blue hexagons bordered in black). The counts have been rescaled from 887.5 MHz to 1.4 GHz by assuming $\alpha=0.7$. Vertical bars represent Poissonian errors on the normalized counts (computed following \citealt{regener51}). Systematic errors due to incompleteness and bias corrections, as well as spectral index assumptions, scale with the size of the plotted symbols. Also shown for comparison are the counts derived from other wide-field ($\gg 1$ deg$^{2}$) surveys. All the surveys shown in this plot are taken at frequencies at or near 1.4 GHz, to minimize systematic errors introduced when rescaling the counts to 1.4 GHz. Unless stated differently, the rescaling has been done by assuming $\alpha=0.7$, as for the G23 survey. The surveys are listed in the legend: these are the FIRST  survey (\citealt{white97}),  the 943.5 MHz EMU Pilot Survey (EMU-PS; \citealt{norris21}), the SDSS STRIPE-82 mosaic (\citealt{heywood16}), the ATESP survey (\citealt{prandoni01}), the Westerbork Lockman Hole mosaic (LHW; \citealt{prandoni18}), the Phoenix Deep Survey (PDF; \citealt{hopkins03}), the 3 GHz JVLA-COSMOS project (rescaled to 1.4 GHz by \citealt{smolcic17b}, by exploiting measured source spectral indices, whenever available), the 610 MHz survey of the ELAIS-N1 (EN1) field (\citealt{ocran20}). Also shown are the recent counts obtained by \citet{matthews21} by combining the NVSS and the MeerKAT DEEP2 field (1 deg$^{2}$), which span eight decades of flux density. The yellow and light blue shaded areas illustrate the predicted cosmic variance effects for survey coverages of 5 and 10 deg$^{2}$ respectively. They have been obtained by splitting the S3-SEX simulation of \citet{wilman08} (covering $1\times 200$ deg$^{2}$, see black line) in forty 5-deg$^{2}$ and twenty 10-deg$^{2}$ fields respectively. The 25 deg$^{2}$ medium tier of the more recent T-RECS simulations (\citealt{bonaldi19}) is represented by the violet shaded area. Finally, the counts derived from the \citet{mancuso17} radio source evolutionary model are indicated by the light green solid  line.}\label{fig:counts}
\end{figure*}

\begin{table}
\centering
\begin{tabular}{cccccc}
\hline \hline
$\Delta S$ & $\langle S\rangle$ & $N_{S}$ & $d$N/$d$S $S^{2.5}$ & $\sigma^-$ & $\sigma^+$ \\

[mJy] & [mJy] &  & ($sr^{-1} Jy^{1.5}$) & &  \\
\hline
0.20  -- 0.29 &  0.24  & 7513  &  7.48   &  0.09   &  0.09 \\          
0.29 -- 0.41  &  0.34  & 8426  &  8.63   &  0.09   &  0.10  \\   
0.41 -- 0.57  &  0.48  & 6148  &  9.01   &  0.11   &  0.12 \\  
0.57 -- 0.81  &  0.68  & 4195  &  9.42   &  0.15    &  0.15 \\ 
0.81 -- 1.15  &  0.97  & 2732  &  9.77   &  0.19   &  0.19  \\ 
1.15 -- 1.63  &  1.37  & 1971  &  11.61  &  0.26   &  0.27   \\     
1.63 -- 2.30  &  1.94  & 1334  &  13.10  &  0.36   &  0.37 \\    
2.30 -- 3.25  &  2.74  & 1020  &  16.77  &  0.53   &  0.54  \\    
3.25 -- 4.60  &  3.87  & 835   &  23.04   &  0.80    &  0.82 \\     
4.60 -- 6.51  &  5.47  & 660   &  30.5   &  1.2    &  1.2 \\         
6.51 -- 9.20  &  7.74  & 551   &  42.9   &  1.8    &  1.9   \\ 
9.20 -- 13.0  &  10.9  & 417   &  54.5   &  2.7   &  2.8 \\    
13.0 -- 26.0  &  18.4  & 627   &  88.1   &  3.5    &   3.7  \\   
26.0 -- 52.1  &  36.8  & 417   &  165.5   &  8.1   &  8.5  \\        
52.1 -- 104  &  73.6  & 213   &  239    &  16   &  17 \\    
104 -- 417   &  208   & 149   &  375   &  31   &  33 \\ 
417 -- 1666  &  833   & 25    &  504   &   101    &  121  \\
\hline    	  
\end{tabular}

\caption{\label{tab-counts} 887.5 MHz source counts as derived from our survey. {\it Column 1:} flux interval ($\Delta$S); {\it Column 2:}  geometric mean of the flux density ($\langle S\rangle$); {\it Column 3:}  number of sources detected ($N_S$); {\it Column 4:}  differential counts normalized to a non evolving Euclidean model (n $S^{2.5}$); {\it Columns 5 and 6:}  Poissonian errors on the normalized counts.}

\end{table}

The differential source counts obtained from the G23 source catalogue, normalised to a non-evolving Euclidean model ($S^{2.5}$), are listed in Table~\ref{tab-counts} and shown in Figure~\ref{fig:counts} (black-bordered blue hexagons). The source counts are corrected for both Eddington bias (\citealt{eddington13,eddington40}) and resolution bias (i.e. the fact that a larger source of a given total flux density will drop below the signal-to-noise threshold of a survey more easily than a smaller source of the same total flux density). This is done following standard recipes in the literature (see e.g. \citealt{prandoni01,prandoni18,mandal21}). The counts are also corrected for catalogue incompleteness and spurious detections, based on the results of the simulations presented in Section~\ref{sec:compl}. 

Figure \ref{fig:counts} compares the G23 source counts (rescaled from 887.5 MHz to 1.4 GHz by assuming $\alpha=0.7$) with counts derived from some of the widest-area samples available to date at 1.4 GHz. This includes sub-mJy surveys covering >1 deg$^2$ regions, like the Phoenix Deep Field
\citep[PDF, 4.6 deg$^2$;][]{hopkins03} and the Westerbork 6 deg$^2$ mosaic covering the Lockman Hole (LH) region (LHW; \citealt{prandoni18}), as well as shallower (> 1 mJy) but larger ($>>10$ sq. degr.) surveys like ATESP \citep[25 deg$^2$;][]{prandoni01}, Sloan Digital Sky Survey (SDSS) Stripe 82 \citep[100 deg$^2$;][]{heywood16} FIRST \citep[1550 deg$^2$;][]{white97} and the very recent determination obtained by combining NVSS and confusion-limited MeerKAT L-band observations of the DEEP2 field \citep[1 deg$^2$;][]{matthews21}. Also shown are counts derived from wide-area sub-mJy surveys taken at similar frequencies to G23: namely the ELAIS-N1 field observed at 610 MHz (EN1, 1.9 deg$^2$; \citealt{ocran20}) and the EMU Pilot Survey (EMU-PS, 270 deg$^2$; \citealt{norris21}), taken at 943.5 MHz. Furthermore we show the deep counts obtained from the 3 GHz JVLA-COSMOS project \citep[2 deg$^2$;][]{smolcic17b}. All such counts are rescaled to 1.4 GHz assuming the same spectral index as for G23, except for the JVLA-COSMOS project, where measured spectral indices were available for a fraction of the sources. Finally we show 1.4 GHz simulated source counts derived from the source evolutionary model of \citet[][light green solid line]{mancuso17}, from the T-RECS 25 deg$^2$ medium tier simulation \citep[][violet shaded area]{bonaldi19}, as well as different realizations obtained from the S3-SEX simulated catalogue \citep{wilman08}: $1\times 200$ deg$^2$ (black solid line), $20\times 10$ deg$^2$ regions (light blue shaded area) and $40\times 5$ deg$^2$ regions (yellow shaded area).

Figure \ref{fig:counts} clearly demonstrates that the source counts derived from the G23 catalogue nicely match previous counts derived from wide-area surveys, and are in good agreement with the most recent simulations \citep{bonaldi19}, at the same time providing very robust statistics over the entire flux range spanned by the source catalogue. It is noteworthy that there is a quite good agreement with the recent EMU-PS counts \citep{norris21}, also obtained with ASKAP, and with \cite{matthews21} counts, which are the most robust  to date, down to 10 $\mu$Jy.

\subsection{Source properties}
\label{sec:sprop}
\begin{figure*}
	\hspace*{-5em}\includegraphics[width=210mm]{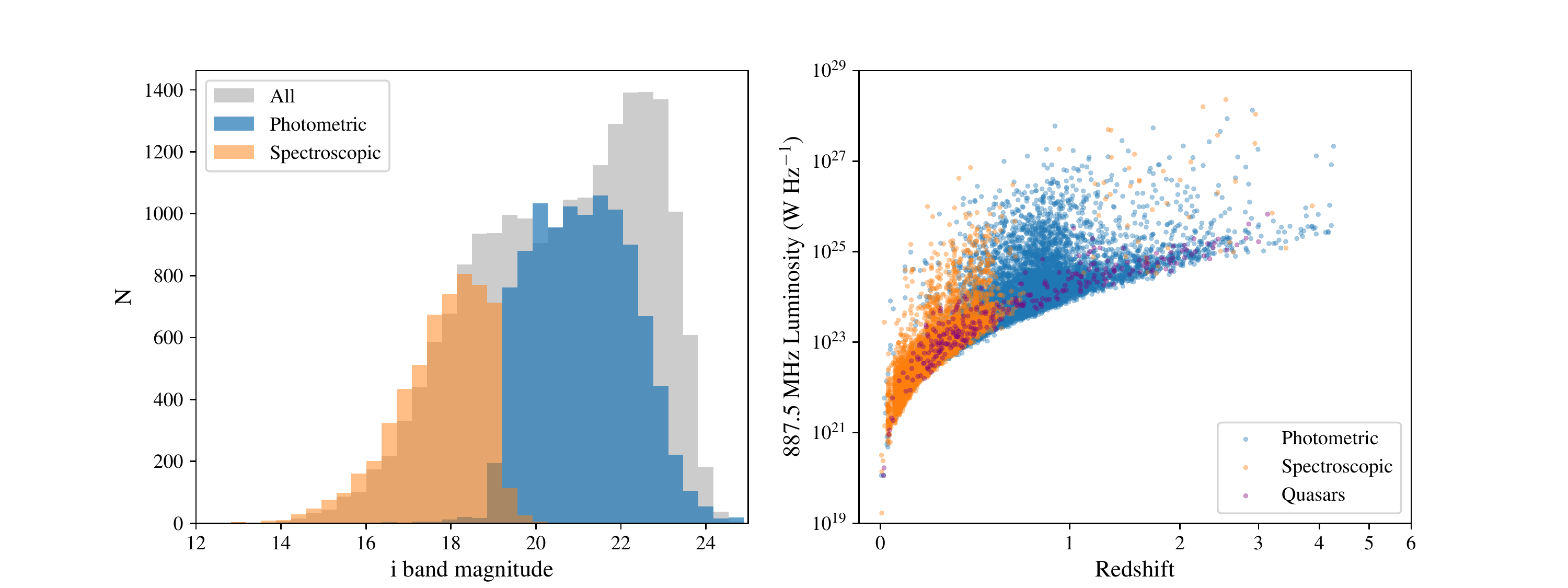}
    \caption{Left: i-band magnitude distribution of identified sources with spectroscopic redshifts, photometric redshifts and all galaxies. Right: Radio luminosity distribution of sources with photometric redshifts, spectroscopic redshifts and quasars as a function of source redshift. Redshift is plotted on a log (1+z) scale but labelled with z values.}
    \label{fig:rband}
\end{figure*}

There are in total 25087 sources with host galaxy counterparts. Out of the 323  multi-component radio sources (identified and combined via visual inspection) 264 have a host galaxy identification, and 56 of these latter have spectroscopic redshifts and 130 have reliable photometric redshifts. We also cross-matched our final catalogue with the latest release of the MilliQuas catalogue which contains quasars and quasar candidates \citep[v7.2;][]{quas21} within 2 arcsec (here we used the coordinates of the radio sources) which provided 591 matches. In the rest of the work we only show sources that are classified as quasars (i.e. the type column equals to Q or q) from this catalogue. In the left panel of Figure \ref{fig:rband} we show the apparent i-band magnitude distribution of sources identified with spectroscopic, photometric and no redshifts. We see that the our radio source sample is virtually spectroscopically complete up to around $i\sim 18$ and becomes severely incomplete at $i> 18.8$. The number of sources with either spectroscopic or photometric redshifts are given in Table \ref{tbl:stat2}.

\begin{table}
\centering
\caption{Number of sources with spectroscopic (spec-z) redshifts and photometric redshifts (photo-z). We also present the number of sources with reliable (see Section \ref{sec-catalogue}) photometric redshifts. }\label{tbl:stat2}
\begin{tabular}{lccc} % four columns, alignment for each
\hline
\hline
Source class & Count\\
\hline
Sources with GAMA counterpart & 20007\\
Sources with WISE counterpart & 5080\\
Sources with spec-z & 5934\\
Sources with photo-z & 14776\\
Sources with reliable photo-z & 11583\\
\hline
\end{tabular}
\end{table}

In the the right panel of the same figure (Fig. \ref{fig:rband}) we show the distribution of rest frame radio luminosity at 887.5 MHz (using a spectral index of 0.7) as a function of source redshift. In this panel we also show sources that have matches in the MilliQuas-V7.2 catalogue (we use our redshifts where available, otherwise we use redshifts published in the MilliQuas v7.2 catalogue for quasars). Similarly it can be seen that the majority of spectroscopic redshifts are limited to z $\sim$ 0.6. The highest redshift object in the sample is a quasar at z=6.44 \citep{Decarli18}. Here we provide a more reliable estimate of this high redshift source using our carefully imaged data. The total flux density of this quasar at 887.5 MHz is 0.68$\pm$0.065 mJy (as opposed to the reported value 0.59$\pm$0.07 mJy by \citet{qso21} using the old ASKAP mosaic of G23) and for a redshift of 6.44 the rest frame radio luminosity of the quasar corresponds to 1.84$\times$10$^{26}$ W/Hz.

\begin{figure*}
	\includegraphics[width=178mm]{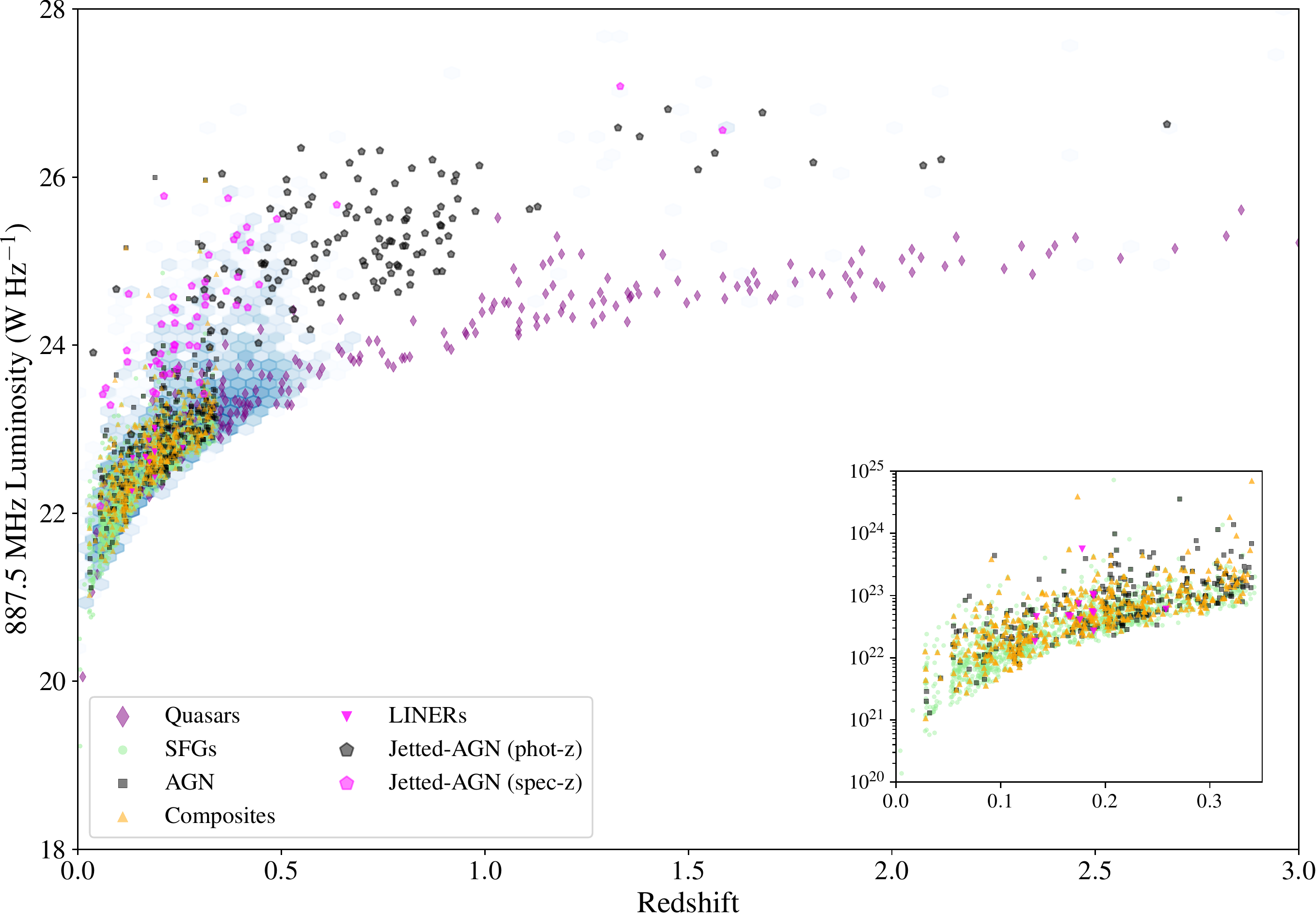}
    \caption{Combined density/scatter figure showing the distribution of the luminosity at 887.5 MHz of all ASKAP-detected galaxies in the sample as a function of their redshifts. Colours and different symbols indicate different emission-line classes. Blue hexagons show the density of sources unclassified by their optical emission lines. Other sources are shown with a symbol for each object: these include SFGs (light green circles) composites (orange triangle point up), AGN (black squares) and LINERs (pink triangle point down) and jetted-AGN (pentagons: magenta with spectroscopic redshift and gray with photometric redshift). Quasars are shown as purple diamonds. The inner panel shows only sources classified by the optical emission lines using the same symbols.}
    \label{fig:classified}
\end{figure*}

We further classify our sources by their optical emission lines available to us \citep{gordon17}. For this we use  Baldwin-Phillips-Terlevich (BPT)-type \citep{bpt} emission-line diagnostics described by \citet{kewley06}. Our classification was carried out using only the following emission lines: [NII]$\lambda$6583,  [SII]$\lambda$6717, H$_{\beta}$, OIII$\lambda$5007, and H$_{\alpha}$. Before implementing our selection we also corrected the flux of H$_{\alpha}$ and H$_{\beta}$ lines for stellar absorption \citep{gordon17}. Composite objects were separated from star-forming galaxies (SFGs) using the criterion given by \citet{kauff03}. It is necessary for objects to have the required optical emission lines – in our case H$_{\beta}$, OIII$\lambda$5007, H$_{\alpha}$, [NII]$\lambda$6583, and [SII]$\lambda$6717 – detected at 3$\sigma$ in order to classify galaxies accurately. The number of sources in each population after optical emission line classification is given in Table \ref{tbl:classified}. The rest frame radio luminosity distribution of sources classified based on the optical emission lines as a function of redshift are seen Fig. \ref{fig:classified}. A large number of galaxies, more than half the parent sample, are not detected in all the required optical emission lines at 3$\sigma$, and those are therefore unclassified by the methods we use. These are shown in Fig. \ref{fig:classified} as blue density plots. As discussed by \citep{gurkan18} a fraction of these sources are a contaminating population of low-luminosity AGN whereas some might be high redshift SFGs. The optically selected sample (AGN: black squares, SFGs: light green circles, Composites: orange triangle point up, LINERs: pink triangle point down) show similar radio luminosities as can be seen in the inner panel of Fig. \ref{fig:classified} except that some AGN have quite high radio luminosity. We also show radio sources having components resolved in our map using pentagon symbols (magenta for objects with spectroscopic redshift and gray for sources with photometric redshift). As expected these sources have much higher radio luminosities than sources selected by their optical emission lines. Quasars (i.e. matches in the MilliQuas-V7.2 catalogue) are shown as purple diamonds. It is not surprising that quasars have the widest redshift coverage and a fraction of these objects present high radio luminosities.

\begin{table}
	\centering
	\caption{The number of sources in each population after the optical emission-line classification and the median spectroscopic redshift for each population.}
	\label{tbl:classified}
	\begin{tabular}{lccccc} % four columns, alignment for each
		\hline
		\hline
		Emission line class & Count & Median redshift\\
		\hline 
	    SFGs&1636&0.163\\
	    Composites&384&0.18\\
	    AGN&261&0.213\\
	    LINERs&13&0.178\\
	    Unclassified&4022&0.227\\
		\hline
	\end{tabular}
\end{table}

\subsection{Giant radio galaxies}
In AGN the energy is generated by accretion of matter onto SMBHs in the centre of active galaxies. This accretion can give rise to relativistic jets, which are mainly observed at radio wavelengths. These AGN driven relativistic jets can have a range of sizes; at typical distances the largest angular scales can reach to several arcminutes, which might be corresponding to hundreds of kpc. Giant radio galaxies (GRGs) have the largest projected linear sizes up to $\sim$5 Mpc \citep{willis74,chandra-saikia99}. \citealt{dabhade20} recently released the largest sample of GRGs identified in a radio survey using data from LoTSS. It is still debated how a fraction of jetted AGN acquire such gigantic sizes. Among the suggested reasons for their large sizes, there a few possibilities considered such as GRGs reside in sparser environments \citep[e.g.][]{saripalli2015,dabhade20}, allowing them to grow fast and reach to larger distances (but see \citealt{lan21}) or GRGs are the result of recurrent radio jet activity \citep{bruni20}. Furthermore, they appear to have different central engine properties \citep{hard19} and lifetime distributions \citep{shabala20} compared to standard jetted AGN.  

\begin{figure*}
	\includegraphics[width=180mm]{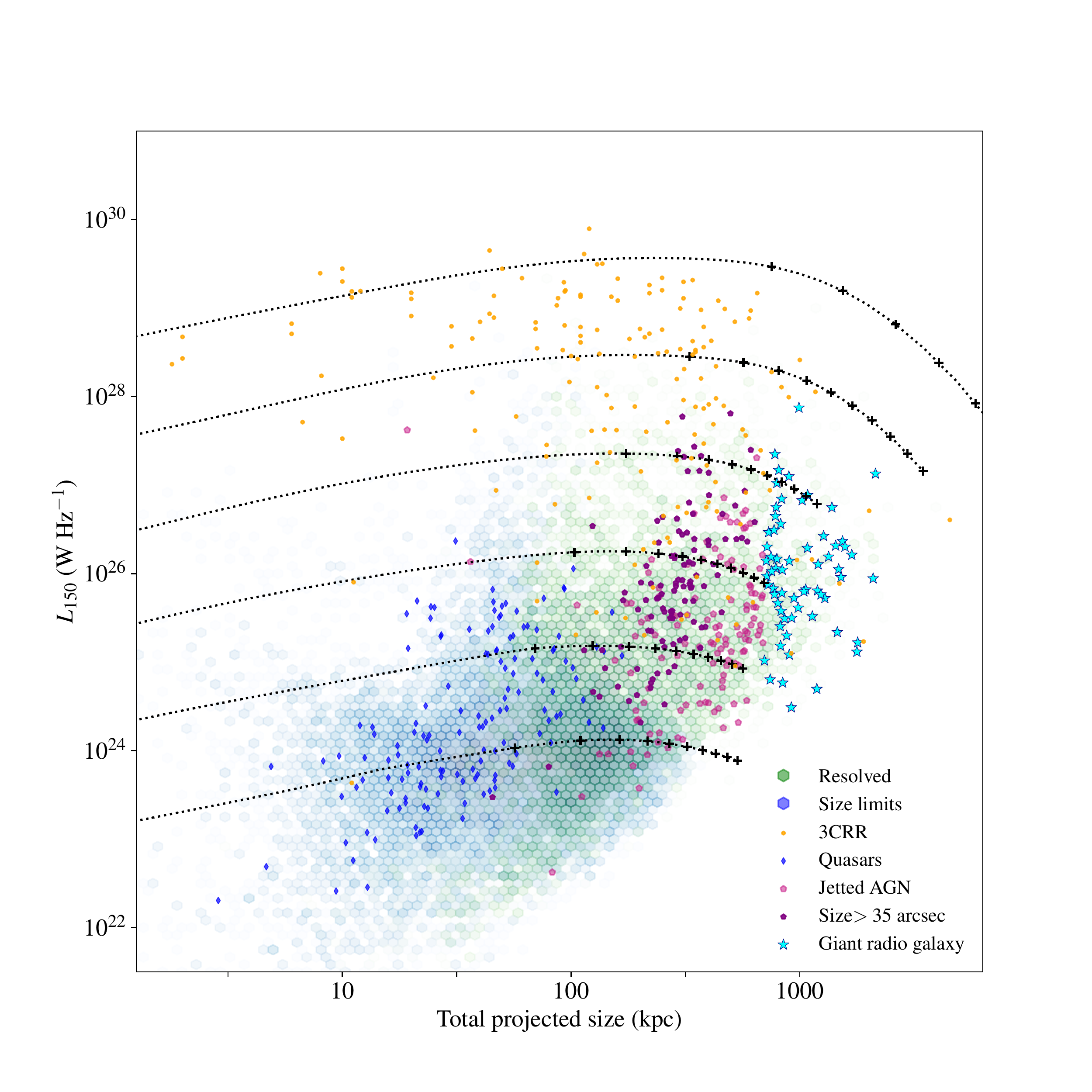}
    \caption{Radio power$-$linear size ($P−D$ diagram) for the AGN sample. Blue and green density hexagons show unresolved and resolved LoTSS jetted-AGN samples \citep{hard19}. Pink circles are 3CRR sources from \citealt{laing83}. Quasars, found in our survey by matching our final catalogue with the MilliQuas v7.2 catalogue \citep{quas21} are shown as blue diamonds. Radio sources with resolved multiple components that were identified through visual inspection are seen as purple pentagons whereas radio sources with sizes larger than 35 arcsec are shown as pink pentagons. Cyan stars indicate GRGs. We also show the theoretical evolutionary tracks published by \citealt{hard19} for $z=0$ sources lying in the plane of the sky in a group environment (M$_{500}$ = 2.5$\times$10$^{13}$ M$_{\odot}$, kT = 1 keV) for two-sided jet powers (from bottom to top) Q = 10$^{35}$,10$^{36}$,...,10$^{40}$ W. Crosses on the tracks are plotted at intervals of 50 Myr, where linear size increases monotonically with time; each track lasts for 500 Myr in total. Here the ASKAP radio luminosities of sources at 887.5 MHz were extrapolated to 150 MHz using a sepctral index of 0.7.}
    \label{fig:pdd}
\end{figure*}

\begin{figure*}
    \includegraphics[width=50mm]{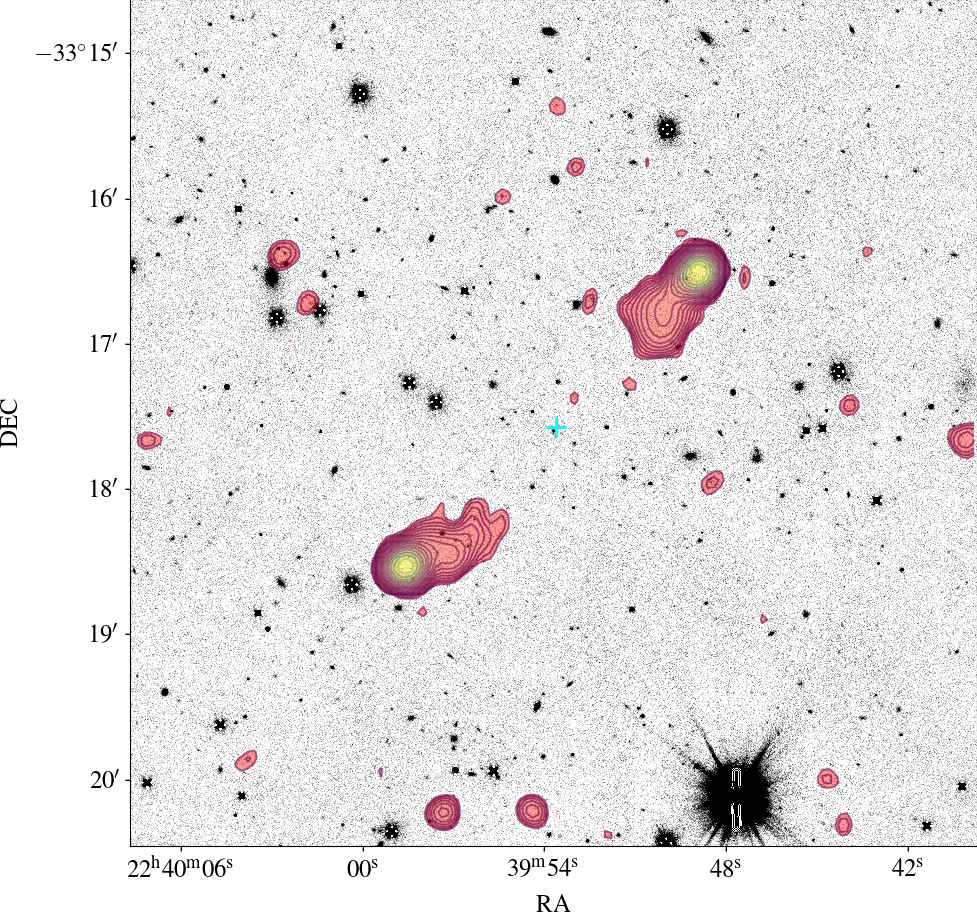}
    \includegraphics[width=50mm]{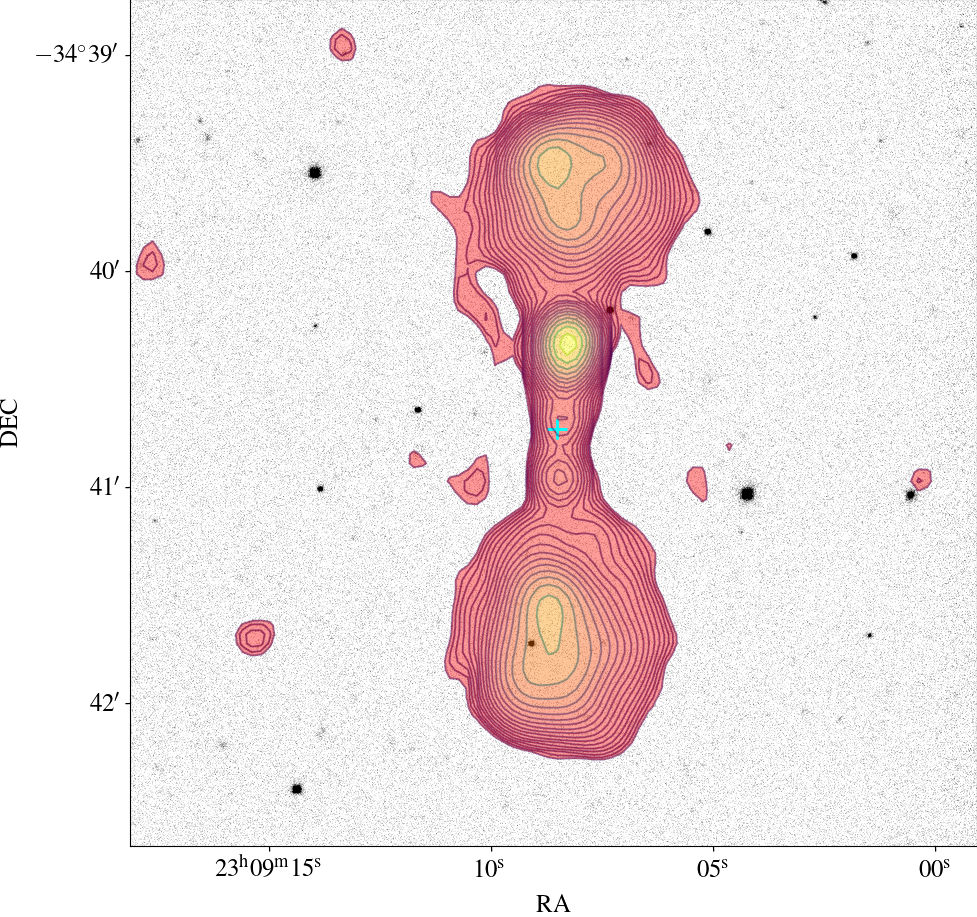}
    \includegraphics[width=50mm]{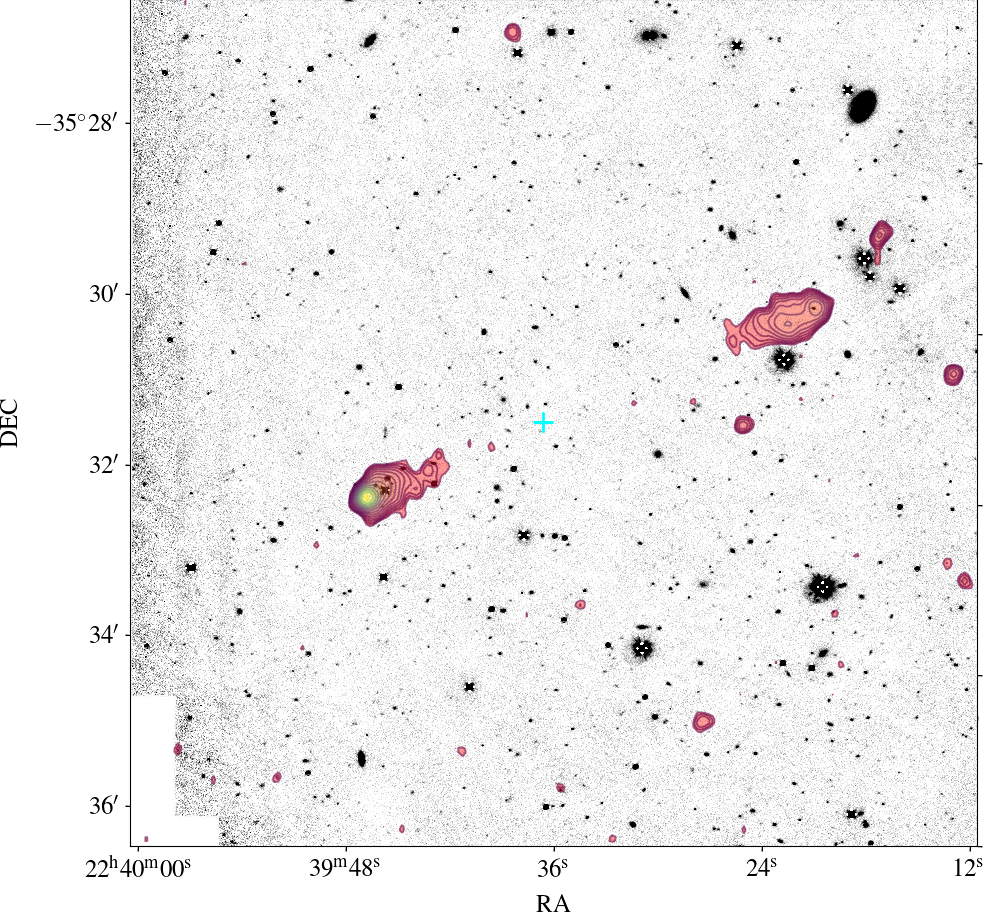}
    \includegraphics[width=50mm]{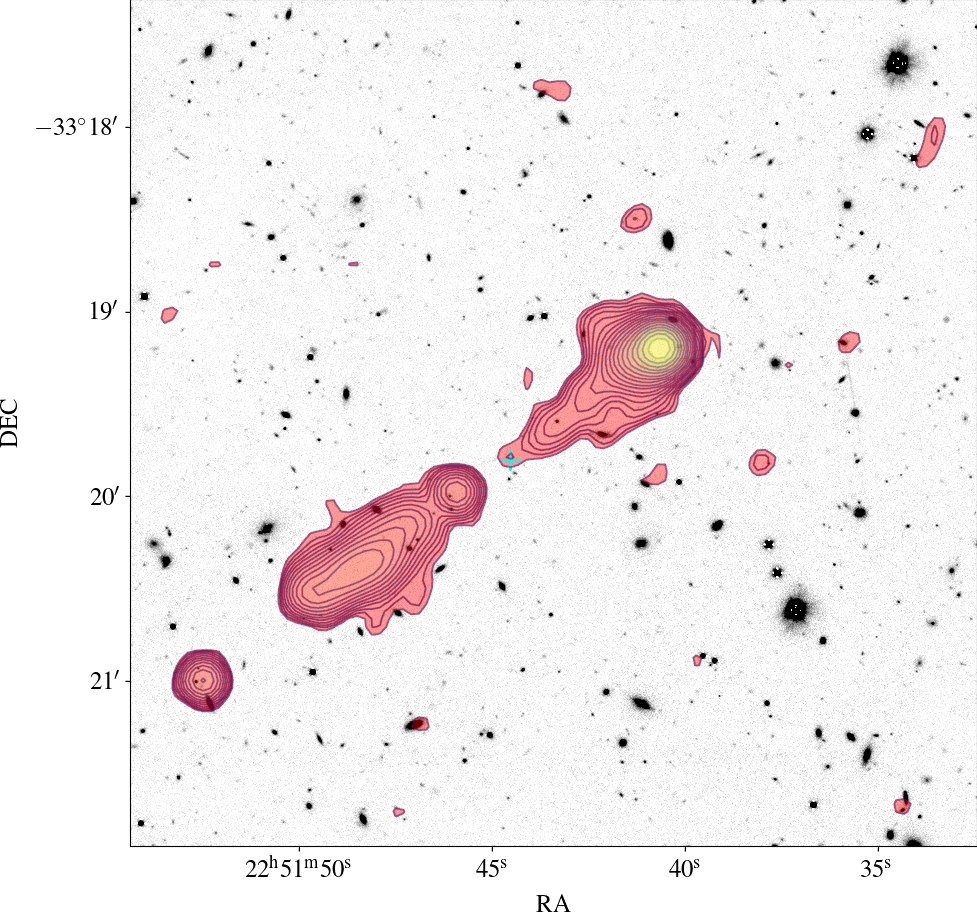}
    \includegraphics[width=50mm]{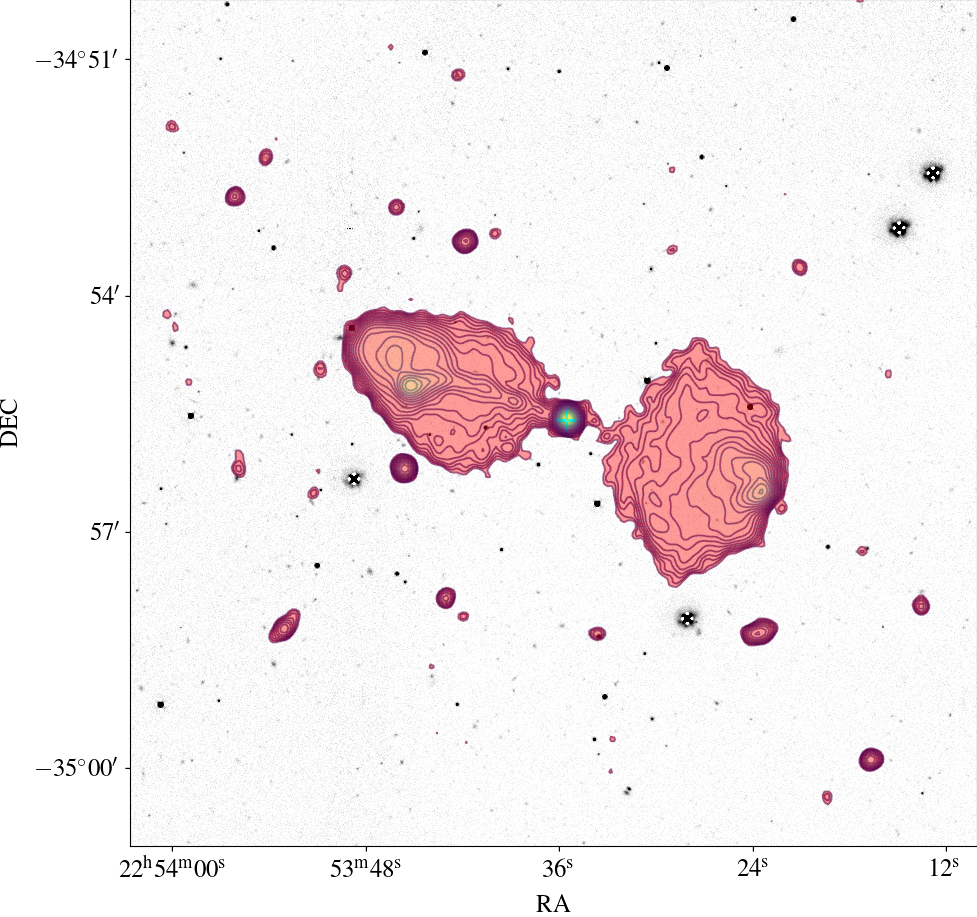}
    \includegraphics[width=50mm]{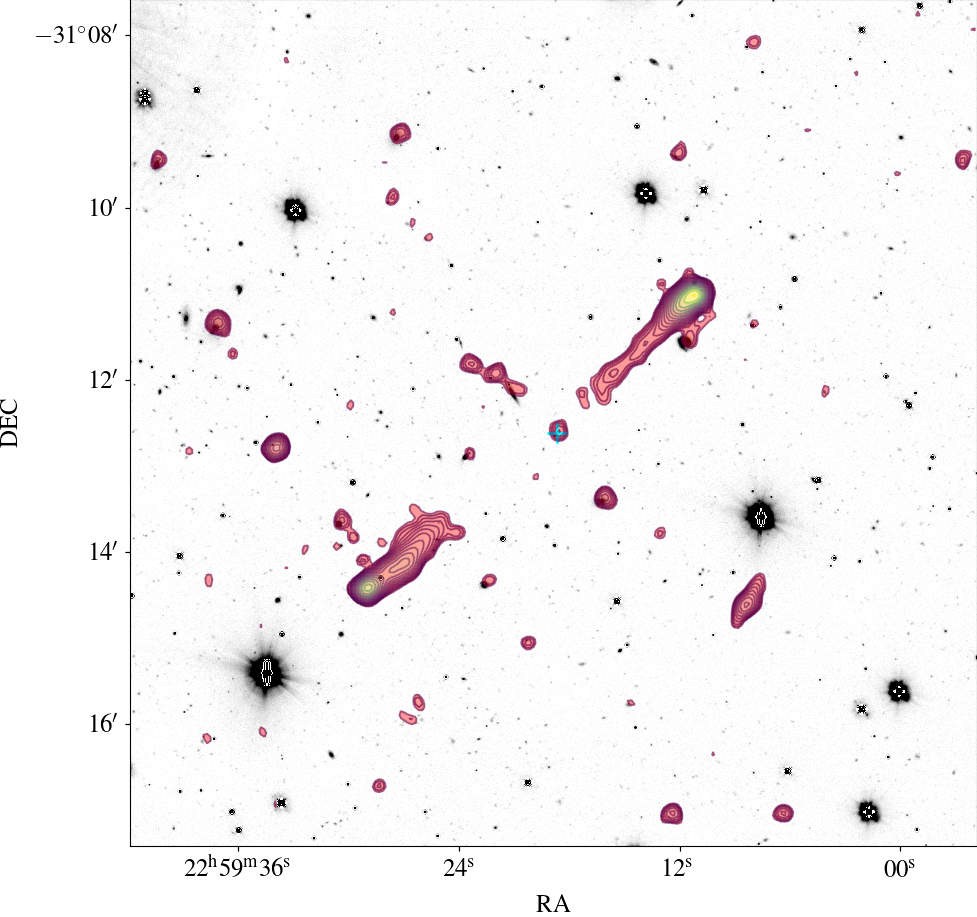}
    \includegraphics[width=50mm]{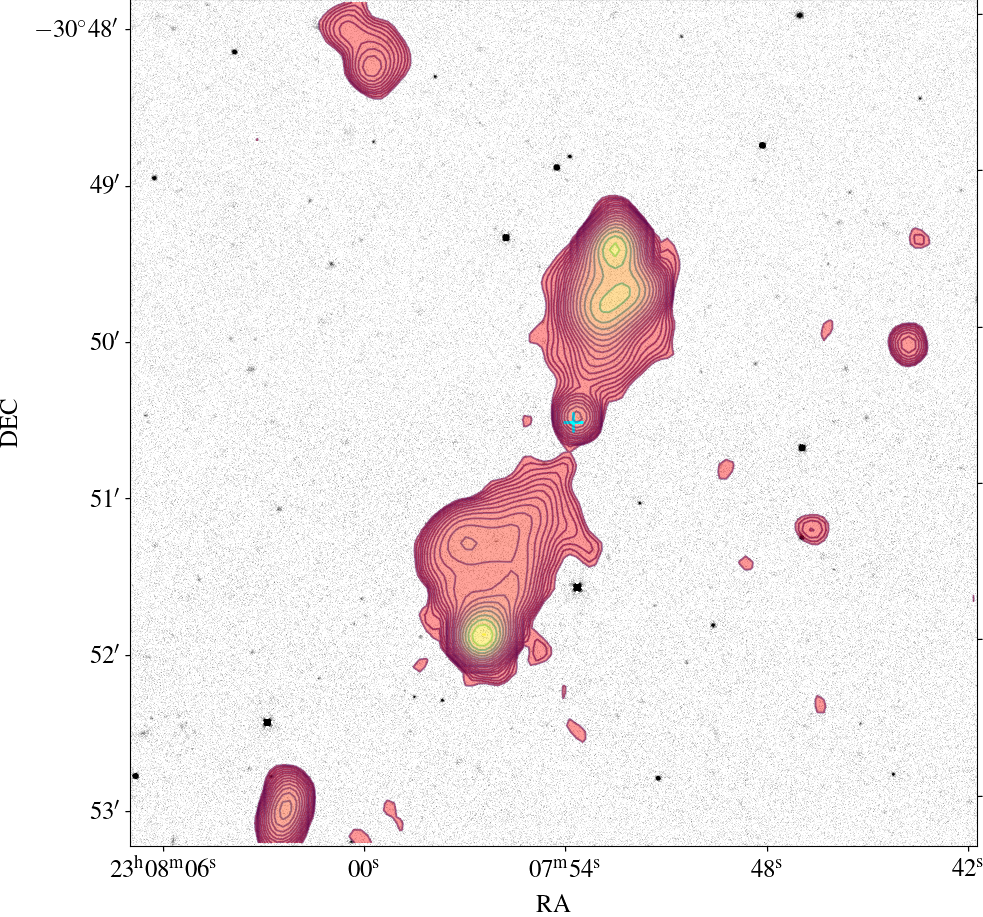}
    \includegraphics[width=50mm]{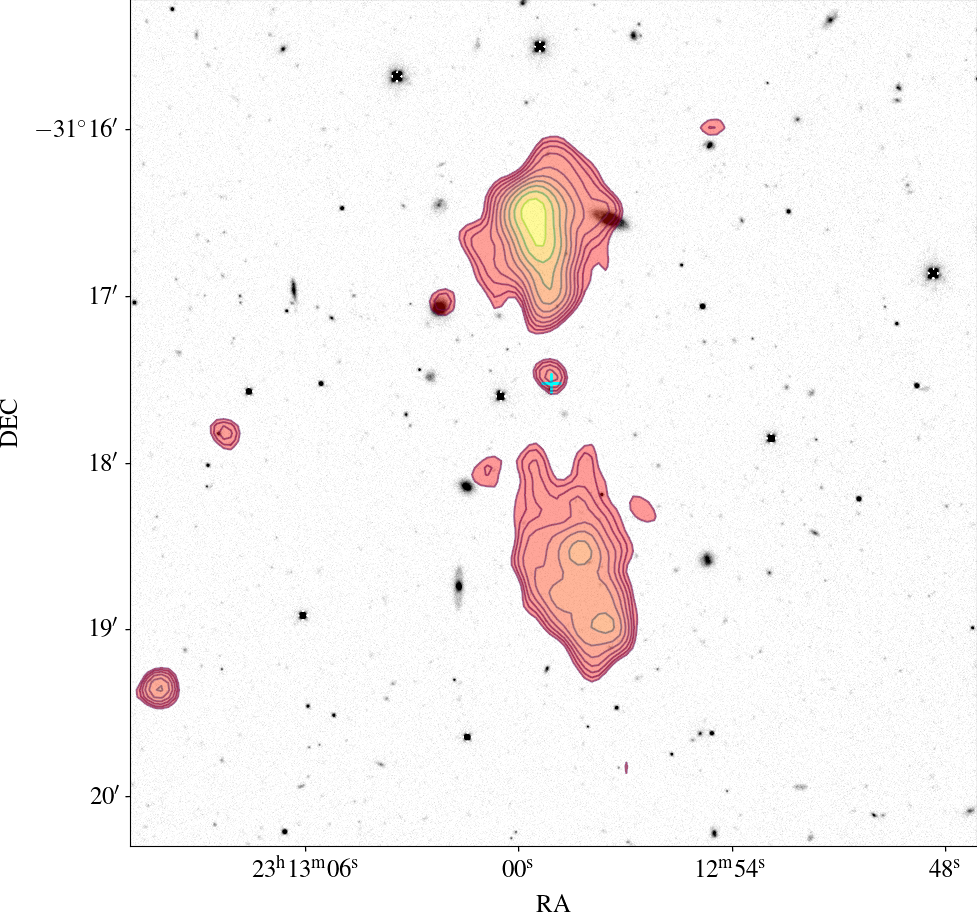}
    \includegraphics[width=50mm]{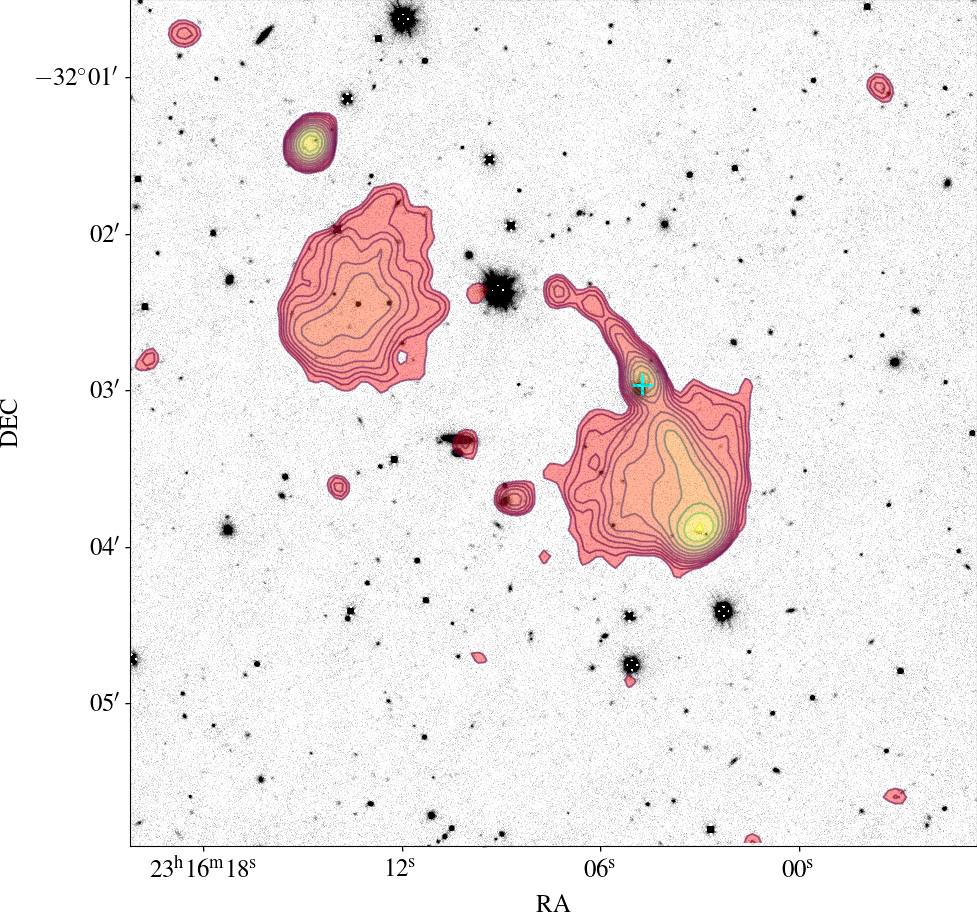}
    \includegraphics[width=50mm]{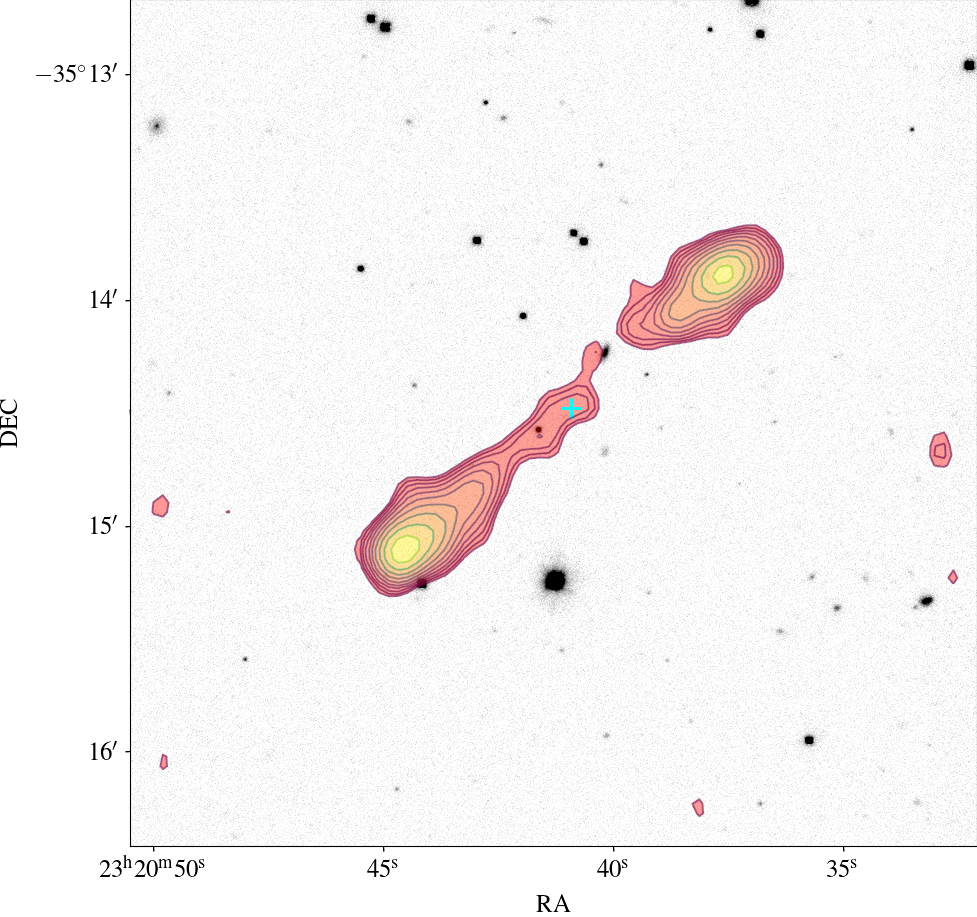}
    \includegraphics[width=50mm]{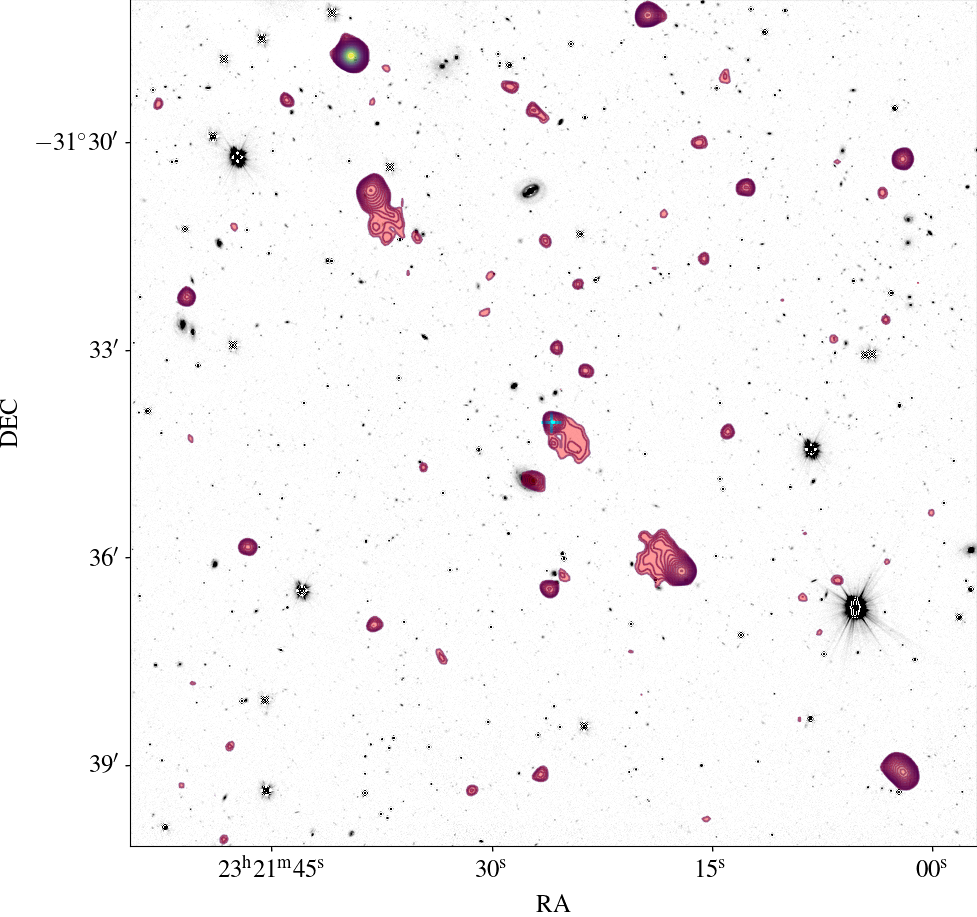}
    \includegraphics[width=55mm]{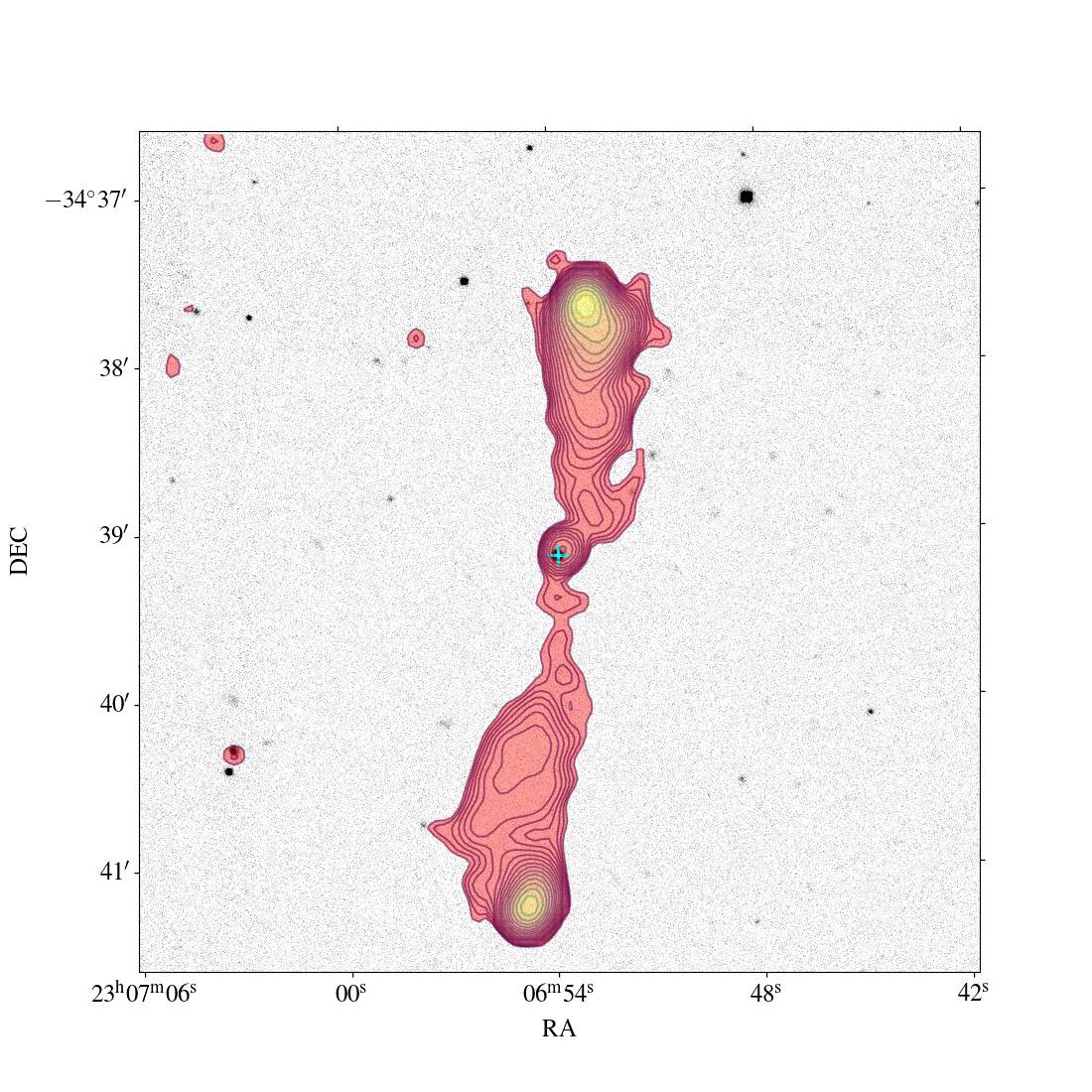}
    \caption{Example of giant radio galaxies newly discovered as part of the ASKAP observations in the G23 field. Background images are from KiDS. Salmon colour density maps show ASKAP data. Cyan crosses show host galaxy positions. Density contours denote the surface brightness levels starting at 3$\sigma$ and increasing at various powers of 3$\sigma$, where $\sigma$ denotes the local RMS noise in each map.}
    \label{fig:grgs}
\end{figure*}

ASKAP's excellent capability of rapidly observing wide fields, providing radio data at high sensitivities and resolution in conjunction with the multi-wavelength data available for the G23 field  allowed us to discover several GRGs, some of which are shown in Fig. \ref{fig:grgs}. These include the well-known GRG studied in detail by \cite[the bottom right one in Fig. \ref{fig:grgs}]{seymour20}. Following the literature we classify a source to be a GRG if its size $>0.7$ Mpc. There are 63 sources (among sources with reliable spectroscopic and photometric redshifts) meeting this criterion. As a sanity check it is interesting to see where our GRGs fall in the so-called P$-$D diagram \citep[i.e. the plane identified by radio power and source linear size of jetted AGN, ][]{baldwin82,turner18,hard18,hard19}. In order to make a comparison to other AGN samples like the 3CRR sample of \citet{laing83} and the one constructed using LoTSS data \citep{hard19}, we extrapolated the radio luminosity of our AGN sample from 887.5 MHz to 150 MHz using a radio spectral index of 0.7). The P$-$D diagram is shown in Fig. \ref{fig:pdd}. Different AGN classes are shown with different symbols (see figure legend).

We also show the theoretical tracks first introduced by \citet{hard18} for z=0 sources lying in the plane of the sky in a group environment (M$_{500}$\footnote{Here M$_{500}$ stands for the total mass of the system within the radius R$_{500}$, kT indicates the temperature of X-ray emitting gas of the galaxy group.} = 2.5$\times$10$^{13}$ M$_{\odot}$, kT = 1 keV) for two-sided jet powers (from bottom to top) Q = 10$^{35}$,10$^{36}$,...,10$^{40}$ W. The $P-D$ diagram has been traditionally used to understand the evolutionary path of jetted-AGN. Although it is not straightforward to infer direct information about AGN based on their location in the $P-D$ diagram (due to various effects such as radio galaxy environment, Doppler boosting and the source angle to the line of sight) it is still invaluable for interpreting evolution of AGN populations and GRGs. The theoretical tracks are derived from a model that predicts the time evolution of both luminosity and physical size in a given environment and for a given jet power Q (defined as the two-sided  power, i.e. the total kinetic power of both jets, for further details we refer the reader to \citealt{hard19}). The positions of the time evolution markers on the tracks show that, if all jetted-AGN have long lifetimes, we expect them to spend most of their lifetime with (unprojected) sizes between a few tens and a few hundreds of kpc. 

In Fig. \ref{fig:pdd} we show resolved AGN that were identified by visual inspection (with clearly resolved components) as well as sources with sizes larger than 35 arcsec (this is because showing only sources identified by visual inspection will miss jetted AGN whose components were combined by the source finder algorithm), quasars and GRGs. If we evaluate our sources displayed in the $P-D$ diagram based on only the normalisation of the tracks we see that the majority of jetted AGN, radio sources with large linear sizes (i.e. size$>35$arcsec) and quasars in these models correspond to jet power $\gtrsim$ 10$^{35}$ W similar to LoTSS AGN, though there is a considerable number of quasars that have jet powers lower than 10$^{35}$ W (occupying the region where we see LoTSS AGN with size limits). The overabundance of these compact sources reinforces the idea that many more short-lived sources occur compared to long-lived ones (i.e. the fact that the P$-$D distribution shows that most sources do $\it{not}$ have sizes of 10s$-$100s kpc), as suggested by \cite{shabala20} and \cite{hard19}, though it should be noted that the radio emission from some of these quasars might be dominated by star formation \citep[e.g.][]{gurkan19,macfarlane21}. Finally, giant radio galaxies that we discovered in our survey occupy the tail in these models and correspond to similar jet powers with extensive sizes ($P>10^{38}$ W, size $>700$ kpc) where we expect jetted-AGN to spend most of their lifetime.

\section{Conclusions and future prospects}\label{sec:5}
We have presented ASKAP observations of the G23 field made as part of the EMU Early Science program. The imaging data products presented as part of this study form the first wide-area (82.7 deg$^{2}$), high-resolution (10 arcsec) and high sensitivity (with a central rms off-source noise of 38$\mu$Jy beam$^{-1}$) survey of the G23 field carried out at 887.5 MHz using ASKAP. We have quantified the positional and flux density accuracy of the ASKAP sources in the field. We have presented the differential source counts at 887.5 MHz, which follow quite well the model predictions of various studies and several radio counts that were published previously using other radio surveys. The outcome of our efforts to combine all physically associated radio sources into single objects (Section \ref{sec-finalcat}) along with their properties (their correct flux densities and sizes) is one of the major products of this paper. Using the multi-wavelength photometric and optical spectroscopic catalogues we identified host galaxy counterparts of the majority of radio sources in the final catalogue (around 63 per cent). The synergy between excellent ASKAP data and multi-wavelength catalogues allowed us to discover new giant radio galaxies in the field (63 in total).  

Subsequent papers will investigate various source classifications and revisit the evaluation of photometric redshifts (Marchetti et al.), explore the link between radio luminosity and AGN outflows for low-luminosity sources (Prandoni et al.) and further investigate the giant radio galaxies (G\"urkan et al.). The data products presented here are expected to be reasonably representative in data quality of the much larger EMU survey currently being carried out. Value-added information constructed as part of this work, assessment of these products as shown here along with the ongoing studies demonstrate excellent capabilities of ASKAP in terms of observing wide fields swiftly, its data quality and the scientific use of its data, which will contribute to the first steps towards the exciting science with ASKAP in the pre-SKA era.

\section*{Acknowledgements}
We thank the anonymous referee for their constructive comments. We thank Aaron Chippendale for assisting us to produce Fig. \ref{fig:tiles}, Daniel Mitchell and Mark Wieringa for their discussion with respect to the mosaicing function and Martin Hardcastle for making the data available to replicate Fig. 8 in their work. IP acknowledges support from INAF under the SKA/CTA PRIN “FORECaST” and the PRIN MAIN STREAM “SAuROS” projects and from CSIRO under its Distinguished Research Visitor Programme. LM, IP and MV acknowledge support from the Italian Ministry of Foreign Affairs and International Cooperation (MAECI Grant Number ZA18GR02) and the South African Department of Science and Innovation's National Research Foundation (DSI-NRF Grant Number 113121) under the ISARP RADIOSKY2020 Joint Research Scheme. DL is supported by a grant from the Natural Sciences and Engineering Research Council of Canada. M. Bilicki is supported by the Polish National Science Center through grants no. 2020/38/E/ST9/00395, 2018/30/E/ST9/00698 and 2018/31/G/ST9/03388, and by the Polish Ministry of Science and Higher
Education through grant DIR/WK/2018/12. This paper includes archived data obtained through the CSIRO ASKAP Science Data Archive, CASDA (\url{http://data.csiro.au}). The Australian SKA Pathfinder is part of the Australia Telescope National Facility (grid.421683.a) which is managed by CSIRO. Operation of ASKAP is funded by the Australian Government with support from the National Collaborative Research Infrastructure Strategy. ASKAP uses the resources of the Pawsey Supercomputing Centre. Establishment of ASKAP, the Murchison Radio-astronomy Observatory and the Pawsey Supercomputing Centre are initiatives of the Australian Government, with support from the Government of Western Australia and the Science and Industry Endowment Fund. We acknowledge the Wajarri Yamatji people as the traditional owners of the Observatory site. We acknowledge the use of the Ilifu cloud computing facility –\url{ www.ilifu.ac.za}, a partnership between the University of Cape Town, the University of the Western Cape, the University of Stellenbosch, Sol Plaatje University, the Cape Peninsula University of Technology and the South African Radio Astronomy Observatory. The Ilifu facility is supported by contributions from the Inter-University Institute for Data Intensive Astronomy (IDIA – a partnership between the University of Cape Town, the University of Pretoria, the University of the Western Cape and the South African Radio astronomy Observatory), the Computational Biology division at UCT and the Data Intensive Research Initiative of South Africa (DIRISA).This research has made use of the University of Hertfordshire high-performance computing facility (\url{http://uhhpc.herts.ac.uk/}) located at the University of Hertfordshire. This research made use of astropy, a community-developed core Python package for astronomy \citep{astropy13} hosted at \url{http://www.astropy.org/} and of TOPCAT \citep{taylor05}. This research has made use of the CIRADA cutout service at \url{cutouts.cirada.ca}, operated by the Canadian Initiative for Radio Astronomy Data Analysis (CIRADA). CIRADA is funded by a grant from the Canada Foundation for Innovation 2017 Innovation Fund (Project 35999), as well as by the Provinces of Ontario, British Columbia, Alberta, Manitoba and Quebec, in collaboration with the National Research Council of Canada, the US National Radio Astronomy Observatory and Australia’s Commonwealth Scientific and Industrial Research Organisation.

%%%%%%%%%%%%%%%%%%%%%%%%%%%%%%%%%%%%%%%%%%%%%%%%%%
\section*{Data Availability}
The data underlying this article are available in the article and in its online supplementary material. Calibrated visibilities of the data (scheduling blocks 8132 and 8137) used for this work are available on the CSIRO ASKAP Science Data Archive (CASDA).

\section*{Supplementary material}
%\setlength{\tabcolsep}{3pt}
%\captionsetup[table]{position=bottom}
%\usepackage[labelfont=bf]{caption}

\subsection*{The associated catalogue}
The final radio source catalogue produced as a result of our analysis and visual inspection  (Section \ref{sec-finalcat}) is made available for download in FITS format as part of the supplementary material. Table \ref{tbl:catalogue} below gives the column names and descriptions.

\begin{table*}
\centering
\caption{The FITS catalogue columns and their description.}\label{tbl:catalogue}
\begin{tabular}{|l|p{0.5\linewidth}}
\hline\hline
Column name (unit) & Description\\
\hline
Source Name &  Source name that are generated the final radio source positions.\\
RA (deg) \& E RA (deg) & Flux-weighted right ascension (RA) and uncertainty.\\
DEC (deg) \& E Dec (deg) & Flux-weighted declination (DEC) and uncertainty.\\
Total flux (Jy) \& E Total Flux (Jy) & Integrated source flux density and uncertainty.\\
Peak flux (Jy beam$^{-1}$) \& E Peak Flux (Jy beam$^{-1}$) & Peak flux density and uncertainty\\
Length (arcsec) & Source length\\
Width (arcsec) & Source width\\
PA (deg) & Position angle\\
DC Maj (arcsec) \& E DC Maj (arcsec) & Deconvolved major axis and uncertainty\\
DC Min (arcsec) \& E DC Min (arcsec) & Deconvolved minor axis and uncertainty\\
Maj (arcsec) \& E Maj (arcsec) & Major axis and uncertainty\\
Min (arcsec) \& E Min (arcsec) & Minor axis and uncertainty\\
Separation (arcsec) & Separation between the farthest components\\
Isl rms (Jy beam$^{-1}$) & The local rms noise used for the source detection\\
SCode & Source morphology code; M indicates sources with multiple components that were identified via visual inspection.\\
Size (arcsec) & Source size (size of multi-component radio sources is estimated as the average of length and separation and deconvolved major axis is used for the rest)\\
\hline
\end{tabular}
\end{table*}


\bibliographystyle{mnras}
\bibliography{askap} % if your bibtex file is called example.bib

%\begin{thebibliography}{99}
%\end{thebibliography}

%%%%%%%%%%%%%%%%%%%%%%%%%%%%%%%%%%%%%%%%%%%%%%%%%%

%%%%%%%%%%%%%%%%% APPENDICES %%%%%%%%%%%%%%%%%%%%%

%\appendix

%%%%%%%%%%%%%%%%%%%%%%%%%%%%%%%%%%%%%%%%%%%%%%%%%%

% Don't change these lines
%\bsp	% typesetting comment
\label{lastpage}
\end{document}